\documentclass[pre,twocolumn]{revtex4}

\usepackage{epsfig}
\usepackage{amssymb,amsmath,stmaryrd,tabularx}
\usepackage{float}
\usepackage{wrapfig}
\usepackage{subfigure}
\usepackage{bm}
\usepackage{graphicx}
\newcommand{\B}[1]{{\bm{#1}}}%% Bold Roman & Greek Lower & Upper Case
\DeclareMathAlphabet{\mathitbf}{OML}{cmm}{b}{it}

\newcommand{\sFrac}[2]{{\textstyle\frac{#1}{#2}}}

\newcommand{\xv}{\mathitbf x}
\newcommand{\Xv}{\mathitbf X}

\newcommand{\fv}{\mathitbf f}
\newcommand{\Hv}{\mathitbf H}
\newcommand{\Iv}{\mathitbf I}
\newcommand{\Av}{\mathitbf A}
\newcommand{\calBold}[1]{\mbox{\boldmath${\cal #1}$}}
\newcommand{\mathBold}[1]{\mbox{\boldmath$#1$}}

\begin{document}
\title{Athermal Nonlinear Elastic Constants of Amorphous Solids}
\author{Smarajit Karmakar, Edan Lerner, and Itamar Procaccia}
\affiliation{Dept of Chemical Physics, The Weizmann Institute of Science, Rehovot 76100, Israel}
\begin{abstract}
We derive expressions for the lowest nonlinear elastic constants of amorphous solids in athermal conditions (up to third order), in terms of
the interaction potential between the constituent particles. The effect of these constants cannot be disregarded when amorphous solids undergo instabilities like plastic flow or fracture in the athermal limit; in such situations
the elastic response increases enormously, bringing the system much beyond the linear regime. We demonstrate that the existing theory of thermal nonlinear elastic constants converges to our expressions in the limit of zero temperature.
We motivate the calculation by discussing two examples in which these nonlinear elastic constants play a crucial role in the context of elasto-plasticity of amorphous solids. The first example is the plasticity-induced memory that is typical
to amorphous solids (giving rise to the Bauschinger effect).  The second example is how to predict the next plastic
event from knowledge of the nonlinear elastic constants. Using the results of this paper we derive a simple
differential equation for the lowest eigenvalue of the Hessian matrix in the external strain near mechanical instabilities; this equation predicts
how the eigenvalue vanishes at the mechanical instability and the value of the strain where the mechanical instability takes place.
\end{abstract}
\date{\today}
\maketitle

%%%%%%%%%%%%%%%%%%%%%%%%%%%%%%%%%%%%%%%%%%%%%%%%%%%
\section{Introduction}
Many phenomena occurring in elastic materials, like plasticity, fracture and shear banding were traditionally
studied assuming that the framework of linear elasticity can be employed to describe the dynamics of the evolving
systems. Recently it became clearer that this is not a good idea; close to the fracture tip, where the stress
field tends to diverge, or near a plastic instability, where the shear modulus diverges towards $-\infty$, nonlinear effects become crucial if not dominant \cite{08ELF,10LBSF,10KLLP}. Moreover, recent studies indicate that nonlinear elastic moduli play an
important role in the memory that amorphous solids exhibit of their loading trajectory \cite{10KLP}. The well known Bauschinger effect
can be explained as a result of the growth of the second order elastic modulus which is identically zero in an isotropic
amorphous solid. It becomes therefore necessary to present a microscopic theory of the nonlinear elastic constants to the aim of computing them in numerical simulations. The present paper has in mind athermal quasistatic simulations, a subject
of high theoretical interest for which there had been great recent progress in understanding the range of phenomena observed, including a variety of scaling laws and the emergence of subsequent scaling theories \cite{04ML,06ML,09LPa,09LPb,09HKLP}. Interestingly, the finite-temperature counterpart of the theory presented in this paper is available in the literature, but it is not quite obvious how to extract from it the athermal limit. We will show below that indeed the thermal theory converges to our
theory when $T\to 0$.

The structure of this paper is as follows. In Sec.~\ref{athermal} we present the derivation of the athermal
elastic constants. In Sec.~\ref{thermal} we present a derivation of the thermal elastic constants; the results
of this derivation are scattered in the literature, but it is worthwhile to present them here in a compact and
consistent notation. In Sec.~\ref{limit} we demonstrate that the limit $T\to 0$ of the thermal elastic constants
reduces properly to our results in Sec.~\ref{athermal}. Finally, in Sec.~\ref{applications} we demonstrate
the usefulness of our results in the context of the plasticity induced memory of amorphous solids and in predicting
mechanical instabilities.

%%%%%%%%%%%%%%%%%%%%%%%%%%%%%%%%%%%%%%%%%%%
\section{Derivation of the athermal nonlinear elastic constants}
\label{athermal}
\subsection{definitions}
We denote the $\nu$ component of the position of the $i$'th particle as $x^i_\nu$;
in the following Roman characters denote particle indices, and Greek characters
denote cartesian components.
Given a linear transformation of coordinates $x^i_\nu \to H_{\nu\kappa}x^i_\kappa$,
the resulting displacement field is $u^i_\nu = H_{\nu\kappa}x^i_\kappa - x^i_\nu$.
The strain tensor is defined up to second order in the derivatives of the displacement field as
\begin{equation}\label{strain}
\epsilon_{\alpha\beta} \equiv \frac{1}{2}
\left(\frac{\partial u_\alpha}{\partial x_\beta} + \frac{\partial u_\beta}{\partial x_\alpha} +
\frac{\partial u_\nu}{\partial x_\alpha}\frac{\partial u_\nu}{\partial x_\beta}\right)\ ,
\end{equation}
where here and below repeated indices are summed over,
unless explicitly indicated otherwise.
In terms of the transformation $\B H$, the strain tensor is
\begin{equation}\label{strainT}
\epsilon_{\alpha\beta} \equiv \sFrac{1}{2}\left(
H^T_{\alpha\nu}H_{\nu\beta} - \delta_{\alpha\beta}\right)\ .
\end{equation}
The displacement field in the definitions (\ref{strain}) and
(\ref{strainT}) should be understood as taken
with respect to the actual configuration, which may be {\em arbitrarily deformed};
the identity of the un-deformed isotropic reference state is of no interest
in the following derivation.

The free energy ${\cal F}$ is defined in terms of the partition function
${\cal Z}$ as ${\cal F} = -T\log{\cal Z}$, where
\begin{equation}
{\cal Z} = \int d\B pd\B q \, e^{-\frac{E}{T}}\ ,
\end{equation}
and $E = U+K$ is the sum of the potential and kinetic energies of the system; $\B p, \B q$ are the momenta and
coordinates of the particles. We expand the free energy density ${\cal F}/V$ up to a constant,
again with respect to the actual configuration, in terms of this strain tensor
\begin{equation}\label{freeEExp}
{\cal F}/V\! =\! \tilde{C}_1^{\alpha\beta}\epsilon_{\alpha\beta} +
\sFrac{1}{2}\tilde{C}_2^{\alpha\beta\nu\eta}\epsilon_{\alpha\beta}\epsilon_{\nu\eta}
+ \sFrac{1}{6}\tilde{C}_3^{\alpha\beta\nu\eta\kappa\chi}
\epsilon_{\alpha\beta}\epsilon_{\nu\eta}\epsilon_{\kappa\chi},
\end{equation}
where
\begin{eqnarray}\label{koo1}
\tilde{C}_1^{\alpha\beta} =  \frac{1}{V}\left. \frac{\partial {\cal F}}
{\partial \epsilon_{\alpha\beta}}\right|_{\epsilon=0} \ &,& \quad
\tilde{C}_2^{\alpha\beta\nu\eta} =
\frac{1}{V}\left.\frac{\partial ^2{\cal F}}{\partial \epsilon_{\alpha\beta}
\partial \epsilon_{\nu\eta}}\right|_{\epsilon=0} \ , \nonumber\\
\tilde{C}_3^{\alpha\beta\nu\eta\kappa\chi} &= & \frac{1}{V}\left.\frac{\partial^3 {\cal F}}
{\partial \epsilon_{\alpha\beta}
\partial \epsilon_{\nu\eta}\partial \epsilon_{\kappa\chi}}\right|_{\epsilon=0}\ .
\end{eqnarray}
In the athermal limit $T\to 0$ the free energy ${\cal F}$ reduces to the potential energy $U$,
then the expansion (\ref{freeEExp}) reads
\begin{eqnarray}\label{energyExp}
&&\lim_{T\to 0}{\cal F}/V = U/V = \\&&C_1^{\alpha\beta}\epsilon_{\alpha\beta}
+ \sFrac{1}{2}C_2^{\alpha\beta\nu\eta}\epsilon_{\alpha\beta}\epsilon_{\nu\eta}
+ \sFrac{1}{6}C_3^{\alpha\beta\nu\eta\kappa\chi}
\epsilon_{\alpha\beta}\epsilon_{\nu\eta}\epsilon_{\kappa\chi}\ , \nonumber
\end{eqnarray}
with the coefficients
\begin{equation}
\begin{split}
C_1^{\alpha\beta} = &\  \lim_{T\to 0}\tilde{C}_1^{\alpha\beta} = \frac{1}{V}
\lim_{T\to 0} \left. \frac{\partial {\cal F}}
{\partial \epsilon_{\alpha\beta}}\right|_{\epsilon=0} \ , \\
C_2^{\alpha\beta\nu\eta} = &\ \lim_{T\to 0}\tilde{C}_2^{\alpha\beta\nu\eta} =
\frac{1}{V}\lim_{T\to 0}\left.\frac{\partial ^2{\cal F}}{\partial \epsilon_{\alpha\beta}
\partial \epsilon_{\nu\eta}}\right|_{\epsilon=0} \ , \\
C_3^{\alpha\beta\nu\eta\kappa\chi} = &\ \lim_{T\to 0}
\tilde{C}_3^{\alpha\beta\nu\eta\kappa\chi} =
\frac{1}{V}\lim_{T\to 0}\left.\frac{\partial^3 {\cal F}}
{\partial \epsilon_{\alpha\beta}
\partial \epsilon_{\nu\eta}\partial \epsilon_{\kappa\chi}}\right|_{\epsilon=0}\ .
\end{split}
\end{equation}
Taking the limit $T \to 0$ of free energy derivatives is equivalent
to taking the same derivatives, but of the {\em potential energy, while satisfying the
constraints} $\fv \equiv -\nabla U = 0$ \cite{89Lutsko}.
In other words, the meaning of constrained derivatives is that
{\em variations with $\epsilon$ keep the net forces at zero}, i.e.
\begin{equation}\label{constraint}
\frac{\partial}{\partial \epsilon} \quad \mbox{such that} \quad
\fv \equiv -\nabla U = 0\ .
\end{equation}
We denote these constrained derivative as $\left. \frac{\partial}{\partial \epsilon}\right|_\fv$,
then
\begin{eqnarray}
C_1^{\alpha\beta} = \frac{1}{V} \left. \frac{\partial U}
{\partial \epsilon_{\alpha\beta}}\right|_\fv \ &,& \quad
C_2^{\alpha\beta\nu\eta} = \frac{1}{V}\left.\frac{\partial ^2{U}}{\partial \epsilon_{\alpha\beta}
\partial \epsilon_{\nu\eta}}\right|_{\fv,\fv} \ , \nonumber \\
C_3^{\alpha\beta\nu\eta\kappa\chi} &=&  \frac{1}{V}\left.\frac{\partial^3 {U}}
{\partial \epsilon_{\alpha\beta}
\partial \epsilon_{\nu\eta}\partial \epsilon_{\kappa\chi}}\right|_{\fv,\fv,\fv}\ ,
\end{eqnarray}
where derivatives in the above expression should be understood as taken
at $\epsilon=0$.
Satisfaction of the constrained derivatives is achieved by
allowing for an additional displacement field $\Xv$ to the imposed one, which we refer to
as the {\em non-affine} displacement field. So, upon imposing a deformation
via some transformation $\Hv$, the total variation in coordinates can
be split into a sum of two terms
\begin{equation}\label{moo8}
x^i_\alpha \to H_{\alpha\beta}x^i_\beta + X^i_\alpha\ ,
\end{equation}
where the $X^i_\alpha$'s are added to the imposed deformation to satisfy the constraint
(\ref{constraint}). The physical meaning of this procedure is that the derivatives are always
computed after minimizing the energy. The original deformation brings the system away from a local
minimum on the potential energy surface and the minimization amounts to an additional non-affine
displacement which must be taken into account. Note that the non-affine field should be found explicitly,
in terms of potential energy derivatives, and this is a part of the derivation below.

The presence of constrained derivatives calls for some care in the calculation. For the benefit of a novice
reader we present in Appendix \ref{consder} some introductory remarks to facilitate further reading.
%%%%%%%%%%%%%%%%%%%%%%%%%%%%%%%%%%%%%%%%%%%%%%%%%%%%%%%%%%%%%%%%%%%%%%%%%%%%%%
\subsection{Non-affine Velocities}
The non-affine
corrections $\Xv$ of Eq.~(\ref{moo8}) play the role of $y$ in equations
(\ref{moo6}-\ref{moo7}), as their evolution is
dictated by the constraint (\ref{constraint}). The role of $x$ in equations
(\ref{moo6}-\ref{moo7}) is played by the strain $\B \epsilon$, such that in complete analogy with (\ref{moo7}),
the constrained partial derivatives with respect to strain
can be written as
\begin{equation}\label{constDer}
\left. \frac{\partial }{\partial \epsilon_{\alpha\beta}} \right|_\fv	 =
\frac{\partial }{\partial \epsilon_{\alpha\beta}} +
\left. \frac{\partial X^j_\nu}{\partial \epsilon_{\alpha\beta}} \right|_\fv
\frac{\partial }{\partial X^j_\nu} \ ,
\end{equation}
where partial derivatives with respect to strain should be understood as taken at constant $\Xv$,
and partial derivatives with respect to coordinates should be understood as taken at constant $\B \epsilon$.
The quantities $\left. \frac{\partial X^j_\nu}{\partial \epsilon_{\alpha\beta}} \right|_\fv$
can be calculated by demanding that the constrained
derivative of the forces vanishes, i.e. $\left. \frac{\partial f^i_\kappa}
{\partial \epsilon_{\alpha\beta}}\right|_{\fv} = 0$. We use (\ref{constDer})
and write
\begin{equation}\label{zoo2}
\left. \frac{\partial f_\kappa^i}{\partial \epsilon_{\alpha\beta}} \right|_{\fv} =
\frac{\partial f_\kappa^i}{\partial \epsilon_{\alpha\beta}}
+ \left. \frac{\partial X_\nu^j}{\partial \epsilon_{\alpha\beta}}\right|_{\fv}
\frac{\partial f_\kappa^i}{\partial X^j_\nu} = 0\ .
\end{equation}
We identify the first term on the RHS of the above relation as the (negative of the) {\em mismatch forces}
$\Xi^i_{\kappa\alpha\beta} \equiv
-\frac{\partial f_\kappa^i}{\partial\epsilon_{\alpha\beta}}
= \frac{\partial^2U}{\partial \epsilon_{\alpha\beta}\partial x^i_\kappa}$,
which can be calculated explicitly from the potential energy, see
Appendix~\ref{potentialEnergyDerivatives} for the case of pairwise potentials.
Notice that derivatives at constant $\B \epsilon$ can be equivalently taken
with respect to the coordinates $x^i_\alpha$ or the non-affine displacements
$X^i_\alpha$, following Eq.~(\ref{moo8}).
The mismatch forces arise due to the imposed affine deformation,
before the action of non-affine correcting displacements. The second term on
the RHS of (\ref{zoo2}) contains the derivative
$\frac{\partial f_\kappa^i}{\partial X^j_\nu} =
-\frac{\partial^2 U}{\partial X^j_\nu\partial X^i_\kappa}
\equiv -{\cal H}^{ij}_{\kappa\nu}$, which is the negative of the Hessian.
Using these definitions of the mismatch forces and the Hessian, Eq.~(\ref{zoo2}) can we rewritten as
\begin{equation}\label{foo5}
\Xi^i_{\kappa\alpha\beta} + {\cal H}^{ij}_{\kappa\nu}
\left. \frac{\partial X^j_\nu}{\partial \epsilon_{\alpha\beta}} \right|_{\fv} = 0\ .
\end{equation}
We now define the non-affine velocities
${\cal V}^i_{\kappa\alpha\beta} \equiv
\left.\frac{\partial X^i_{\kappa}}{\partial \epsilon_{\alpha\beta}}\right|_{\fv}$,
which can be calculated by inverting Eq.~(\ref{foo5}):
\begin{equation}\label{nonAffVel}
{\cal V}^i_{\kappa\alpha\beta} = \left.\frac{\partial X^i_\kappa}{\partial \epsilon_{\alpha\beta}}\right|_{\fv}
= - ({\cal H}^{-1})^{ij}_{\kappa\nu}\Xi^j_{\nu\alpha\beta}\ .
\end{equation}
Notice that the Hessian is not generally invertible due to zero-modes
that arise from the translational invariance of the potential energy; to overcome this
we expand the RHS of (\ref{nonAffVel}) in eigenfunctions of the Hessian
\begin{equation}\label{ModeExp}
{\cal V}^i_{\kappa\alpha\beta} =
- \sum_{j,\ell}\frac{(\psi_\ell)_\nu^j\Xi^j_{\nu\alpha\beta}}{\lambda_\ell}
(\psi_\ell)^i_\kappa\ ,
\end{equation}
where $(\psi_\ell)_\nu^j$ is the $\nu$ component of the $j$'th particle contribution to
the $\ell$'th eigenfunction, $\lambda_\ell$ is the corresponding eigenvalue,
and the sum over $\ell$ should {\bf not} include the zero-modes.

With the definition of the non-affine velocities,
the constrained partial derivative (\ref{constDer}) reads
\begin{equation}\label{constDer2}
\left. \frac{\partial }{\partial \epsilon_{\alpha\beta}} \right|_\fv	 =
\frac{\partial }{\partial \epsilon_{\alpha\beta}} +
{\cal V}^i_{\nu\alpha\beta}
\frac{\partial }{\partial X^i_\nu}\ ,
\end{equation}
which is the form that will be used from this point on.
%%%%%%%%%%%%%%%%%%%%%%%%%%%%%%%%%%%%%%%%%%%%%%%%%%%%%%%%%%%%%%%%%%%%%%%%%%%%%%%%%%%%%%%
\subsection{First order elastic constants}
The first order athermal elastic constants are
\begin{equation}\label{stressExpansion}
C_1^{\alpha\beta} \equiv \left.
\!\sFrac{1}{V}\sFrac{\partial U}{\partial \epsilon_{\alpha\beta}}\right|_{\fv} \ .
\end{equation}
Equation (\ref{moo8}) implies that forces can be calculated by taking derivatives
with respect to either set of coordinates, i.e.
\begin{equation}\label{moo9}
\frac{\partial U}{\partial X^i_\nu}
= \frac{\partial U}{\partial x^i_\nu}\ ,
\end{equation}
which can be used in calculating the constrained derivative
\begin{equation}
\left.\frac{\partial U}{\partial \epsilon_{\alpha\beta}}\right|_{\fv}=
\frac{\partial U}{\partial \epsilon_{\alpha\beta}} +
{\cal V}^i_{\nu\alpha\beta}
\frac{\partial U}{\partial X^i_\nu}
=\frac{\partial U}{\partial \epsilon_{\alpha\beta}}
-{\cal V}^i_{\nu\alpha\beta} f^i_\nu\ .
\end{equation}
The second term on the RHS of the above equation vanishes due to the
constraint $\fv = 0$ as appears in (\ref{constraint}),
and we are left with
\begin{equation}\label{moo1}
\left.\frac{\partial U}{\partial \epsilon_{\alpha\beta}}\right|_{\fv} =
\frac{\partial U}{\partial \epsilon_{\alpha\beta}}\ .
\end{equation}
Finally,
\begin{equation}\label{firstOrderAthermal}
C_1^{\alpha\beta} =
\frac{1}{V} \frac{\partial U}{\partial \epsilon_{\alpha\beta}}\ .
\end{equation}
Notice that the first order elastic constants
{\bf do not} contain a relaxation term; Noticing the stress $\B \sigma = \B C_1$, this means that
the non affine displacement field does not relax the stress.
For an explicit expression of the potential energy derivatives
in the case of pairwise potentials, see Appendix~\ref{potentialEnergyDerivatives}.

%%%%%%%%%%%%%%%%%%%%%%%%%%%%%%%%%%%%%%%%%%%%%%%%%%%%%%%%%%%%%%%%%%%%%%%%
\subsection{Second order elastic constants}
The second order athermal elastic constants $C_2^{\alpha\beta\nu\eta}$ are
given by $\frac{1}{V}\left. \frac{\partial^2 U}
{\partial \epsilon_{\nu\eta}\partial \epsilon_{\alpha\beta}} \right|_{\fv,\fv}$,
evaluated at $\epsilon = 0$; first, we write
\[
\left. \frac{\partial^2 U}
{\partial \epsilon_{\nu\eta}\partial \epsilon_{\alpha\beta}} \right|_{\fv,\fv}
= \left. \frac{\partial }
{\partial \epsilon_{\nu\eta}} \right|_{\fv}
\left. \frac{\partial U}{\partial \epsilon_{\alpha\beta}} \right|_\fv
= \left. \frac{\partial }
{\partial \epsilon_{\nu\eta}} \right|_{\fv}
\frac{\partial U}{\partial \epsilon_{\alpha\beta}}\ ,
\]
where the second equality follows from (\ref{moo1}). We use the rule
(\ref{constDer2}) and calculate
\begin{equation}\label{secondDerivativeAtConstantF}
\begin{split}
\left. \frac{\partial }
{\partial \epsilon_{\nu\eta}} \right|_{\fv}
\frac{\partial U}{\partial \epsilon_{\alpha\beta}} = &
\left( \frac{\partial }{\partial \epsilon_{\nu\eta}} +
{\cal V}^i_{\theta\nu\eta}
\frac{\partial }{\partial X^i_\theta}
\right) \frac{\partial U}{\partial \epsilon_{\alpha\beta}} \\
= & \frac{\partial ^2 U}{\partial \epsilon_{\nu\eta}
\partial \epsilon_{\alpha\beta}}
+ {\cal V}^i_{\theta\nu\eta}
\frac{\partial^2U}{\partial X^i_\theta \partial \epsilon_{\alpha\beta}}\\
= & \frac{\partial ^2 U}{\partial \epsilon_{\nu\eta}
\partial \epsilon_{\alpha\beta}}
+ {\cal V}^i_{\theta\nu\eta}\Xi^i_{\theta\alpha\beta}\ .
\end{split}
\end{equation}
The quantities $\frac{\partial ^2 U}{\partial \epsilon_{\nu\eta}
\partial \epsilon_{\alpha\beta}}$ and
$\Xi^i_{\nu\alpha\beta} =  \frac{\partial^2U}
{\partial X^i_\nu \partial \epsilon_{\alpha\beta}}$
can be directly calculated from the potential, see Appendix~\ref{potentialEnergyDerivatives}
for the case of pairwise potentials. Finally
\begin{equation}\label{secondOrderElasticConstants}
C_2^{\alpha\beta\nu\eta} = \frac{1}{V}\left(
\frac{\partial ^2 U}{\partial \epsilon_{\nu\eta}
\partial \epsilon_{\alpha\beta}}
+ {\cal V}^i_{\theta\nu\eta}\Xi^i_{\theta\alpha\beta}\right)\ .
\end{equation}

%%%%%%%%%%%%%%%%%%%%%%%%%%%%%%%%%%%%%%%%%%%%%%%%%%%%%%%%%%%%%%%%%%%%%%%%%%%%%%%%%%%%
\subsection{Third order elastic constants}
The third order elastic constants $C_3^{\alpha\beta\nu\eta\kappa\chi}$ are
given by $\frac{1}{V}\left. \frac{\partial^3 U}
{\partial \epsilon_{\kappa\chi}\partial
\epsilon_{\nu\eta}\partial \epsilon_{\alpha\beta}}
\right|_{\fv,\fv,\fv}$,
evaluated at $\epsilon = 0$;
We carry out the constrained derivative of (\ref{secondDerivativeAtConstantF})
\begin{widetext}
\begin{equation}\label{foo4}
\begin{split}
\left. \frac{\partial^3 U}
{\partial \epsilon_{\kappa\chi}\partial \epsilon_{\nu\eta}\partial \epsilon_{\alpha\beta}}
\right|_{\fv,\fv,\fv}
\!\!\!\!= & \left( \frac{\partial }{\partial \epsilon_{\kappa\chi}} +
{\cal V}^i_{\theta\kappa\chi}
\frac{\partial }{\partial X^i_\theta}
\right) \!\! \left( \frac{\partial ^2 U}{\partial \epsilon_{\nu\eta}
\partial \epsilon_{\alpha\beta}} +
{\cal V}^j_{\zeta\nu\eta}\Xi^j_{\zeta\alpha\beta} \right) \\
= & \frac{\partial ^3 U}{\partial \epsilon_{\kappa\chi}\partial \epsilon_{\nu\eta}
\partial \epsilon_{\alpha\beta}} +
 \frac{\partial {\cal V}^j_{\zeta\nu\eta}}{\partial \epsilon_{\kappa\chi}}
\Xi^j_{\zeta\alpha\beta} +
 {\cal V}^j_{\zeta\nu\eta}\frac{\partial \Xi^j_{\zeta\alpha\beta} }
{\partial \epsilon_{\kappa\chi}}\\
& \hspace{-0.2cm} + {\cal V}^i_{\theta\kappa\chi}
\frac{\partial^3 U}{\partial X^i_\theta\partial \epsilon_{\nu\eta}
\partial \epsilon_{\alpha\beta} }
+ {\cal V}^i_{\theta\kappa\chi}\frac{\partial {\cal V}^j_{\zeta\nu\eta}}
{\partial X^i_\theta}\Xi^j_{\zeta\alpha\beta}
+ {\cal V}^i_{\theta\kappa\chi}{\cal V}^j_{\zeta\nu\eta}
\frac{\partial \Xi^j_{\zeta\alpha\beta}}{\partial X^i_\theta}.
\end{split}
\end{equation}
We notice now that according to (\ref{constDer2})
\begin{equation}
\frac{\partial {\cal V}^j_{\zeta\nu\eta}}{\partial \epsilon_{\kappa\chi}}
\Xi^j_{\zeta\alpha\beta} + {\cal V}^i_{\theta\kappa\chi}
\frac{\partial {\cal V}^j_{\zeta\nu\eta}}
{\partial X^i_\theta}\Xi^j_{\zeta\alpha\beta} =
\left. \frac{\partial {\cal V}^j_{\zeta\nu\eta}}{\partial \epsilon_{\kappa\chi}}\right|_{\fv}
\Xi^j_{\zeta\alpha\beta}\ .
\end{equation}
Also $\frac{\partial^3 U}{\partial X^i_\theta\partial \epsilon_{\nu\eta}
\partial \epsilon_{\alpha\beta} }
= \frac{\partial \Xi^i_{\theta\alpha\beta}}{\partial \epsilon_{\nu\eta}}
= \frac{\partial \Xi^i_{\theta\nu\eta}}{\partial \epsilon_{\alpha\beta}}$; with
these relations, (\ref{foo4}) becomes
\begin{equation}\label{thirdConstrainedDerivativeU}
\left. \frac{\partial^3 U}
{\partial \epsilon_{\kappa\chi}\partial \epsilon_{\nu\eta}\partial \epsilon_{\alpha\beta}}
\right|_{\fv,\fv,\fv}
\!\!= \frac{\partial ^3 U}{\partial \epsilon_{\kappa\chi}\partial \epsilon_{\nu\eta}
\partial \epsilon_{\alpha\beta}} +
\left. \frac{\partial {\cal V}^j_{\zeta\nu\eta}}{\partial \epsilon_{\kappa\chi}}\right|_{\fv}
\Xi^j_{\zeta\alpha\beta} +
{\cal V}^j_{\zeta\nu\eta}\frac{\partial \Xi^j_{\zeta\alpha\beta} }
{\partial \epsilon_{\kappa\chi}}
 \hspace{-0.2cm} + {\cal V}^i_{\theta\kappa\chi}
\frac{\partial \Xi^i_{\theta\nu\eta}}{\partial \epsilon_{\alpha\beta}}
+ {\cal V}^i_{\theta\kappa\chi}{\cal V}^j_{\zeta\nu\eta}
\frac{\partial \Xi^j_{\zeta\alpha\beta}}{\partial X^i_\theta}.
\end{equation}
The above expression requires knowledge of the quantity
$\left. \frac{\partial {\cal V}^j_{\zeta\nu\eta}}{\partial \epsilon_{\kappa\chi}}\right|_{\fv}$,
for which an explicit expression is unavailable, but
can be obtained by demanding $-\left. \frac{\partial }{\partial \epsilon_{\kappa\chi}}\right|_\fv
\left.\frac{\partial f^\ell_\nu}{\partial \epsilon_{\alpha\beta}}\right|_{\fv} = 0$:
\begin{equation}
-\left. \frac{\partial }{\partial \epsilon_{\kappa\chi}}\right|_\fv
\left.\frac{\partial f^m_\nu}{\partial \epsilon_{\alpha\beta}}\right|_{\fv} =
\left. \frac{\partial }{\partial \epsilon_{\kappa\chi}}\right|_\fv\left(
\Xi^m_{\nu\alpha\beta} + {\cal H}^{m j}_{\nu\eta}{\cal V}^j_{\eta\alpha\beta}\right)
=\left.\frac{\partial \Xi^m_{\nu\alpha\beta}}{\partial \epsilon_{\kappa\chi}}\right|_{\fv}
+ \left. \frac{\partial {\cal H}^{m j}_{\nu\eta}}{\partial \epsilon_{\kappa\chi}}\right|_{\fv}
{\cal V}_{\eta\alpha\beta}^j \ +
{\cal H}^{m j}_{\nu\eta}\left.
\frac{\partial {\cal V}^j_{\eta\alpha\beta}}{\partial \epsilon_{\kappa\chi}}\right|_\fv = 0\ .
\label{27}
\end{equation}
Now,
\[
\left.\frac{\partial \Xi^m_{\nu\alpha\beta}}{\partial \epsilon_{\kappa\chi}}\right|_{\fv} =
\frac{\partial \Xi^m_{\nu\alpha\beta}}{\partial \epsilon_{\kappa\chi}}
+ {\cal V}^i_{\theta\kappa\chi}
\frac{\partial \Xi^m_{\nu\alpha\beta}}{\partial X^i_\theta}\ , \quad \mbox{and} \quad
\left.\frac{\partial {\cal H}^{m j}_{\nu\eta}}{\partial \epsilon_{\kappa\chi}}\right|_{\fv} =
\frac{\partial {\cal H}^{m j}_{\nu\eta}}{\partial \epsilon_{\kappa\chi}}
+ {\cal V}^i_{\theta\kappa\chi}
\frac{\partial {\cal H}^{m j}_{\nu\eta}}{\partial X^i_\theta}\ .
\]
With these Eq.~(\ref{27}) becomes
\begin{equation}\label{foo1}
\frac{\partial \Xi^m_{\nu\alpha\beta}}{\partial \epsilon_{\kappa\chi}}
+ {\cal V}^i_{\theta\kappa\chi}
\frac{\partial \Xi^m_{\nu\alpha\beta}}{\partial X^i_\theta} +
\frac{\partial {\cal H}^{m j}_{\nu\eta}}{\partial \epsilon_{\kappa\chi}}
{\cal V}_{\eta\alpha\beta}^j
+ {\cal V}^i_{\theta\kappa\chi}
\frac{\partial {\cal H}^{m j}_{\nu\eta}}{\partial X^i_\theta}
{\cal V}_{\eta\alpha\beta}^j+
{\cal H}^{m j}_{\nu\eta}\left.
\frac{\partial {\cal V}^j_{\eta\alpha\beta}}{\partial \epsilon_{\kappa\chi}}\right|_\fv = 0\ .
\end{equation}
With the identity $\frac{\partial {\cal H}^{m j}_{\nu\eta}}{\partial \epsilon_{\kappa\chi}}
= \frac{\partial^3 U}{\partial \epsilon_{\kappa\chi}
\partial X_\nu^m\partial X_\eta^j}
= \frac{\partial \Xi^m_{\nu\kappa\chi}}{\partial X^j_\eta}
=  \frac{\partial \Xi^j_{\eta\kappa\chi}}{\partial X^m_\nu}$,
we invert (\ref{foo1}) to get an expression for $\left. \frac{\partial {\cal V}^j_{\eta\alpha\beta}}
{\partial \epsilon_{\kappa\chi}}\right|_{\fv}$
\begin{equation}\label{nonAffineAcc}
\left. \frac{\partial {\cal V}^j_{\eta\alpha\beta}}
{\partial \epsilon_{\kappa\chi}}\right|_{\fv} =
-({\cal H}^{-1})^{jm}_{\eta\nu}
\left( \frac{\partial \Xi^m_{\nu\alpha\beta}}
{\partial \epsilon_{\kappa\chi}}
+{\cal V}^i_{\theta\kappa\chi}
\frac{\partial \Xi^m_{\nu\alpha\beta}}{\partial X^i_\theta}
+{\cal V}_{\theta\alpha\beta}^i
\frac{\partial \Xi^m_{\nu\kappa\chi}}{\partial X^i_\theta}
+{\cal V}^i_{\theta\kappa\chi}
\frac{\partial {\cal H}^{m q}_{\nu\zeta}}{\partial X^i_\theta}
{\cal V}_{\zeta\alpha\beta}^q  \right) \ .
\end{equation}
Inserting this relation back in (\ref{thirdConstrainedDerivativeU}), together with
(\ref{nonAffVel}) we arrive at the expression for the athermal third order
elastic constants
\begin{equation}\label{thirdOrderAthermalDerivative}\begin{split}
C_3^{\alpha\beta\nu\eta\kappa\chi}
 = \frac{1}{V}&\left[\frac{\partial ^3 U}{\partial \epsilon_{\kappa\chi}\partial \epsilon_{\nu\eta}
\partial \epsilon_{\alpha\beta}} +
{\cal V}^i_{\theta\alpha\beta}{\cal V}^j_{\zeta\nu\eta}
{\cal V}^\ell_{\rho\kappa\chi}\frac{\partial^3 U}
{\partial X^i_\theta \partial X^j_\zeta \partial X^\ell_\rho} \right. \\
&\  + {\cal V}^i_{\theta\alpha\beta}{\cal V}^j_{\zeta\nu\eta}
\frac{\partial \Xi^j_{\zeta\kappa\chi}}{\partial X^i_\theta}
+ {\cal V}^i_{\theta\nu\eta}{\cal V}^j_{\zeta\kappa\chi}
\frac{\partial \Xi^j_{\zeta\alpha\beta}}{\partial X^i_\theta}
+ {\cal V}^i_{\theta\kappa\chi}{\cal V}^j_{\zeta\nu\eta}
\frac{\partial \Xi^j_{\zeta\alpha\beta}}{\partial X^i_\theta}
\\ &\  +{\cal V}^i_{\zeta\alpha\beta}\frac{\partial \Xi^i_{\zeta\nu\eta}}
{\partial \epsilon_{\kappa\chi}} +
{\cal V}^i_{\zeta\nu\eta}\frac{\partial \Xi^i_{\zeta\kappa\chi}}
{\partial \epsilon_{\alpha\beta}} +
\left. {\cal V}^i_{\zeta\kappa\chi}\frac{\partial \Xi^i_{\zeta\alpha\beta}}
{\partial \epsilon_{\nu\eta}} \right]\ .
\end{split}\end{equation}
\end{widetext}

%%%%%%%%%%%%%%%%%%%%%%%%%%%%%%%%%%%%%%%%%%%%%%%%%%%%%%%%%%%%%%%%%%%%%%%%%%%%%%%%%%%%%%%%%%%%
\section{Thermal Elastic Constants}
\label{thermal}

The derivation of the thermal elastic constants, in contrast with their athermal counterpart, can be found
in the literature, see for example \cite{89Lutsko} for the first and second order objects. We provide below
also the third order constants.

The elastic constants at finite temperatures are given by derivatives
of the free energy with respect to strain, see Eq.~\ref{koo1}. We first expand the total energy up to third order in the strain
\begin{widetext}
\begin{equation}
\label{totalEnEx}
E = E_0 + E_1^{\alpha\beta}\epsilon_{\alpha\beta} +
\sFrac{1}{2}E_2^{\alpha\beta\nu\eta}\epsilon_{\alpha\beta}\epsilon_{\nu\eta}
+\sFrac{1}{6}E_3^{\alpha \beta \nu\eta \kappa \chi}\epsilon_{\alpha\beta}
\epsilon_{\nu\eta}\epsilon_{\kappa\chi}\ ,
\end{equation}
with the definitions
\begin{equation}\label{E_terms}
\begin{split}
E_0 & \equiv  E(\B \epsilon=0) \ , \\
E_1^{\alpha\beta} & \equiv  \left.\frac{\partial E}{\partial\epsilon_{\alpha\beta}}
\right|_{\B \epsilon=0}
 =  \left.\frac{\partial K}{\partial\epsilon_{\alpha\beta}}\right|_{\B \epsilon=0} +
 \left.\frac{\partial U}{\partial\epsilon_{\alpha\beta}}\right|_{\B \epsilon=0} \ , \\
E_2^{\alpha\beta\nu\eta} & \equiv
 \left.\frac{\partial^2 E}{\partial\epsilon_{\nu\eta}
 \partial\epsilon_{\alpha\beta}}\right|_{\B \epsilon=0} =
 \left.\frac{\partial^2 K}{\partial\epsilon_{\nu\eta}
 \partial\epsilon_{\alpha\beta}}\right|_{\B \epsilon=0}
 + \left.\frac{\partial^2 U}{\partial\epsilon_{\nu\eta}
 \partial\epsilon_{\alpha\beta}}\right|_{\B \epsilon=0} \ , \\
E_3^{\alpha \beta \nu \eta \kappa \chi} & \equiv \left.\frac{\partial^3 E}
{\partial\epsilon_{\kappa\chi}
\partial\epsilon_{\nu\eta}\partial\epsilon_{\alpha\beta}}\right|_{\B \epsilon=0} =
\left.\frac{\partial^3 K}
{\partial\epsilon_{\kappa\chi}
\partial\epsilon_{\nu\eta}\partial\epsilon_{\alpha\beta}}\right|_{\B \epsilon=0}
+\left.\frac{\partial^3 U}
{\partial\epsilon_{\kappa\chi}
\partial\epsilon_{\nu\eta}\partial\epsilon_{\alpha\beta}}\right|_{\B \epsilon=0}\ .
\end{split}
\end{equation}
Explicit expressions for kinetic energy derivatives can be found in
Appendix~\ref{Kinender}, and potential energy derivatives for
the case of pairwise potentials are available in Appendix~\ref{potentialEnergyDerivatives}.
We expand the Boltzmann factor
\begin{equation}\label{BoltzmannExpansion}
\begin{split}
e^{-\frac{E}{T}}  \simeq &\  e^{-\frac{E_0}{T}}
\hspace{-0.1cm}\left[1 - \frac{E_1^{\alpha\beta}}{T}\epsilon_{\alpha\beta}
+ \frac{1}{2}\left( \frac{E_1^{\alpha\beta}E_1^{\nu\eta}}{T^2} -
\frac{E_2^{\alpha\beta\nu\eta}}{T} \right)
\epsilon_{\alpha\beta}\epsilon_{\nu\eta} \right. \\
& \quad -\frac{1}{6} \left(
\frac{E_1^{\alpha\beta}E_1^{\nu\eta}E_1^{\kappa\chi}}{T^3}
-\frac{E_1^{\alpha\beta}E_2^{\nu\eta\kappa\chi}}{T^2}
-\frac{E_1^{\nu\eta}E_2^{\alpha\beta\kappa\chi}}{T^2}
-\frac{E_1^{\kappa\chi}E_2^{\alpha\beta\nu\eta}}{T^2}
+\left.\frac{E_3^{\alpha\beta\nu\eta\kappa\chi}}{T}\right)
\epsilon_{\alpha\beta}\epsilon_{\nu\eta}\epsilon_{\kappa\chi} \right] .
\end{split}
\end{equation}
The partition function can now be written as
\begin{equation}\label{partitionFunctionExpansion}
{\cal Z} = {\cal Z}_0 + {\cal Z}_1^{\alpha\beta}\epsilon_{\alpha\beta} +
\sFrac{1}{2}{\cal Z}_2^{\alpha\beta\nu\eta}\epsilon_{\alpha\beta}\epsilon_{\nu\eta}
+ \sFrac{1}{6}{\cal Z}_3^{\alpha\beta\nu\eta\kappa\chi}
\epsilon_{\alpha\beta}\epsilon_{\nu\eta}\epsilon_{\kappa\chi}\ ,
\end{equation}
where
\begin{equation}
\begin{split}
{\cal Z}_0  \equiv {\cal Z}(\B \epsilon=0)
 & =  \int e^{-\frac{E_0}{T}}d\B qd\B p\ , \\
{\cal Z}_1^{\alpha\beta} \equiv
\left.\frac{\partial {\cal Z}}{\partial \epsilon_{\alpha\beta}}
\right|_{\B \epsilon=0}&
= -\int e^{-\frac{E_0}{T}} \frac{E_1^{\alpha\beta}}{T}\,d\B qd\B p\ , \\
{\cal Z}_2^{\alpha\beta\nu\eta} \equiv
\left.\frac{\partial^2 {\cal Z}}{\partial \epsilon_{\nu\eta}\partial \epsilon_{\alpha\beta}}
\right|_{\B \epsilon=0}
& = \int e^{-\frac{E_0}{T}}
\left( \frac{E_1^{\alpha\beta}E_1^{\nu\eta}}{T^2} -
\frac{E_2^{\alpha\beta\nu\eta}}{T} \right)d\B qd\B p\ , \\
{\cal Z}_3^{\alpha\beta\nu\eta\kappa\chi} \equiv
\left.\frac{\partial^3 {\cal Z}}{\partial \epsilon_{\kappa\chi}\partial \epsilon_{\nu\eta}\partial \epsilon_{\alpha\beta}}
\right|_{\B \epsilon=0}& =
-\int e^{-\frac{E_0}{T}}
\left(\frac{E_1^{\alpha\beta}E_1^{\nu\eta}E_1^{\kappa\chi}}{T^3}
+\frac{E_3^{\alpha\beta\nu\eta\kappa\chi}}{T} \right. \\
&\quad\quad\quad\quad\quad\quad \left. -\frac{E_1^{\alpha\beta}E_2^{\nu\eta\kappa\chi}}{T^2}
-\frac{E_1^{\nu\eta}E_2^{\alpha\beta\kappa\chi}}{T^2}
-\frac{E_1^{\kappa\chi}E_2^{\alpha\beta\nu\eta}}{T^2}\right)d\B qd\B p\ .
\end{split}
\end{equation}
With these definitions, we extract the following relations
\begin{equation}\label{relations}\begin{split}
\frac{{\cal Z}_1^{\alpha\beta}}{{\cal Z}_0} & = -\frac{1}{T}\langle E_1^{\alpha\beta} \rangle \ ,\\
\frac{{\cal Z}_2^{\alpha\beta\nu\eta}}{{\cal Z}_0} & = \frac{1}{T^2}
\langle E_1^{\alpha\beta}E_1^{\nu\eta} \rangle - \frac{1}{T}
\langle E_2^{\alpha\beta\nu\eta}\rangle \ ,\\
\frac{{\cal Z}_3^{\alpha\beta\nu\eta\kappa\chi}}{{\cal Z}_0} & = -\frac{1}{T^3}
\langle E_1^{\alpha\beta}E_1^{\nu\eta}E_1^{\kappa\chi} \rangle
-\frac{1}{T}\langle E_3^{\alpha\beta\nu\eta\kappa\chi}\rangle \\
&\quad\quad + \frac{1}{T^2}\left(
\langle E_1^{\alpha\beta}E_2^{\nu\eta\kappa\chi}\rangle
+\langle E_1^{\nu\eta}E_2^{\alpha\beta\kappa\chi}\rangle
+\langle E_1^{\kappa\chi}E_2^{\alpha\beta\nu\eta}\rangle \right) \ ,
\end{split}\end{equation}
where triangular brackets denote equilibrium averaging.
\end{widetext}

\subsection{First order thermal elastic constants}
The first order free energy derivative with respect to strain is
\begin{equation}
\frac{\partial {\cal F}}
{\partial \epsilon_{\alpha\beta}} = -T\frac{1}{{\cal Z}}
\frac{\partial {\cal Z}}{\partial\epsilon_{\alpha\beta}}\ .
\end{equation}
With the expansion (\ref{partitionFunctionExpansion}) and relations (\ref{relations}),
this derivative at $\B \epsilon=0$ is
\begin{equation}\label{firstOrderDerivative}
\left.\frac{\partial {\cal F}}{\partial \epsilon_{\alpha\beta}}\right|_{\B \epsilon=0}
= -\frac{T}{{\cal Z}_0}{\cal Z}_1^{\alpha\beta} =  \langle E_1^{\alpha\beta} \rangle \ .
 \end{equation}
The first order elastic constants are thus given by
\begin{equation}
\tilde{C}_1^{\alpha\beta}  = \sFrac{1}{V}\langle E_1^{\alpha\beta} \rangle\ .
\end{equation}

\subsection{Second order thermal elastic constants}
The second order free energy derivative with respect to strain is
\begin{equation}\label{foo3}
\frac{\partial ^2{\cal F}}{ \partial \epsilon_{\nu\eta}
\partial\epsilon_{\alpha\beta}}
 = -T\left(\frac{1}{{\cal Z}}\frac{\partial^2{\cal Z}}
{\partial \epsilon_{\nu\eta}\partial\epsilon_{\alpha\beta}}
-\frac{1}{{\cal Z}^2}
\frac{\partial{\cal Z}}
{\partial \epsilon_{\nu\eta}}\frac{\partial{\cal Z}}
{\partial \epsilon_{\alpha\beta}}\right).
\end{equation}
With the expansion (\ref{partitionFunctionExpansion}) and relations (\ref{relations}),
this derivative at $\mathBold{\epsilon}=0$ is
\begin{eqnarray}\label{secondOrderDerivative}
\left. \frac{\partial ^2{\cal F}}{ \partial \epsilon_{\nu\eta}
\partial\epsilon_{\alpha\beta}}\right|_{\B \epsilon=0}\!\!\! & = &\
T\left(\frac{{\cal Z}_1^{\alpha\beta}{\cal Z}_1^{\nu\eta}}{{\cal Z}_0^2} -
\frac{{\cal Z}_2^{\alpha\beta\nu\eta}}{{\cal Z}_0}\right)  \nonumber \\
&=& \frac{\langle E_1^{\alpha\beta}\rangle
\langle E_1^{\nu\eta}\rangle \!-\! \langle E_1^{\alpha\beta}E_1^{\nu\eta}\rangle}{T}
+ \langle E_2^{\alpha\beta\nu\eta}\rangle \nonumber \\
& = & \ \langle E_2^{\alpha\beta\nu\eta}\rangle
-\sFrac{1}{T}\langle \Delta E_1^{\alpha\beta}\Delta E_1^{\nu\eta}\rangle \ ,
\end{eqnarray}
where $\Delta A \equiv A -
\langle A\rangle$ for any quantity $A$.
The second order elastic constants are
thus given by
\begin{equation}
\tilde{C}_2^{\alpha\beta\nu\eta} = \frac{1}{V}\left[
\langle E_2^{\alpha\beta\nu\eta}\rangle
-\frac{1}{T}\langle \Delta E_1^{\alpha\beta}\Delta E_1^{\nu\eta}\rangle\right]\ .
\end{equation}

\subsection{Third order thermal elastic constants}
The third order free energy derivative with respect to strain is
\begin{widetext}
\begin{equation}\begin{split}
\frac{\partial ^3{\cal F}}{\partial \epsilon_{\kappa\chi} \partial \epsilon_{\nu\eta}
\partial\epsilon_{\alpha\beta}} = \
 -T & \left(
-\frac{1}{{\cal Z}^2}\frac{\partial{\cal Z}}{\partial \epsilon_{\kappa\chi}}
\frac{\partial^2{\cal Z}}{\partial \epsilon_{\nu\eta}\partial\epsilon_{\alpha\beta}}
+ \frac{1}{{\cal Z}}\frac{\partial ^3{\cal Z}}
{\partial \epsilon_{\kappa\chi} \partial \epsilon_{\nu\eta}
\partial\epsilon_{\alpha\beta}} \right. \\
&\ \left. \ +\frac{2}{{\cal Z}^3}\frac{\partial{\cal Z}}{\partial \epsilon_{\kappa\chi}}
\frac{\partial{\cal Z}}
{\partial \epsilon_{\nu\eta}}\frac{\partial{\cal Z}}
{\partial \epsilon_{\alpha\beta}} - \frac{1}{{\cal Z}^2}
\frac{\partial{\cal Z}}{\partial \epsilon_{\alpha\beta}}
\frac{\partial^2{\cal Z}}{\partial \epsilon_{\kappa\chi}\partial\epsilon_{\nu\eta}}
- \frac{1}{{\cal Z}^2}
\frac{\partial{\cal Z}}{\partial \epsilon_{\nu\eta}}
\frac{\partial^2{\cal Z}}{\partial \epsilon_{\kappa\chi}\partial\epsilon_{\alpha\beta}}
\right) \ .
\end{split}\end{equation}
With the expansion (\ref{partitionFunctionExpansion}) and relations (\ref{relations}),
this derivative at $\B \epsilon=0$ is
\begin{equation}\label{thirdOrderDerivative}\begin{split}
\left.\frac{\partial ^3{\cal F}}{\partial \epsilon_{\kappa\chi} \partial \epsilon_{\nu\eta}
\partial\epsilon_{\alpha\beta}}\right|_{\B \epsilon=0} = &\ -T\left(
\frac{{\cal Z}_1^{\alpha\beta}}{{\cal Z}_0}\frac{{\cal Z}_2^{\nu\eta\kappa\chi}}{{\cal Z}_0} +
\frac{{\cal Z}_1^{\nu\eta}}{{\cal Z}_0}\frac{{\cal Z}_2^{\alpha\beta\kappa\chi}}{{\cal Z}_0} +
\frac{{\cal Z}_1^{\kappa\chi}}{{\cal Z}_0}\frac{{\cal Z}_2^{\alpha\beta\nu\eta}}{{\cal Z}_0}
+ \frac{{\cal Z}_3^{\alpha\beta\nu\eta\kappa\chi}}{{\cal Z}_0}
+ 2 \frac{{\cal Z}_1^{\alpha\beta}}{{\cal Z}_0}
\frac{{\cal Z}_1^{\nu\eta}}{{\cal Z}_0}
\frac{{\cal Z}_1^{\kappa\chi}}{{\cal Z}_0}\right) \\
= &\ \langle E_3^{\alpha\beta\nu\eta\kappa\chi}\rangle
- \frac{\langle E_1^{\alpha\beta}E_2^{\nu\eta\kappa\chi}\rangle }{T}
-\frac{\langle E_1^{\nu\eta}E_2^{\alpha\beta\kappa\chi}\rangle }{T}
-\frac{\langle E_1^{\kappa\chi}E_2^{\alpha\beta\nu\eta}\rangle }{T}\\
&\
+\frac{\langle E_1^{\alpha\beta}E_1^{\nu\eta}E_1^{\kappa\chi} \rangle}{T^2}
+2\frac{\langle E_1^{\alpha\beta}\rangle\langle E_1^{\nu\eta}\rangle\langle E_1^{\kappa\chi}\rangle}{T^2}
 + \langle E_1^{\alpha\beta}\rangle\left(\frac{\langle E_1^{\nu\eta}E_1^{\kappa\chi}\rangle}{T^2}
-\frac{\langle E_2^{\nu\eta\kappa\chi}\rangle}{T}\right) \\
&\
 + \langle E_1^{\nu\eta}\rangle\left(\frac{\langle E_1^{\alpha\beta}E_1^{\kappa\chi}\rangle}{T^2}
-\frac{\langle E_2^{\alpha\beta\kappa\chi}\rangle}{T}\right)
 + \langle E_1^{\kappa\chi}\rangle\left(\frac{\langle E_1^{\alpha\beta}E_1^{\nu\eta}\rangle}{T^2}
-\frac{\langle E_2^{\alpha\beta\nu\eta}\rangle}{T}\right) \\
= &\ \langle E_3^{\alpha\beta\nu\eta\kappa\chi}\rangle
+\frac{\langle \Delta E_1^{\alpha\beta}\Delta E_1^{\nu\eta}\Delta E_1^{\kappa\chi} \rangle}{T^2} \\
&\quad - \frac{\langle \Delta E_1^{\alpha\beta}\Delta E_2^{\nu\eta\kappa\chi}\rangle }{T}
- \frac{\langle \Delta E_1^{\nu\eta}\Delta E_2^{\alpha\beta\kappa\chi}\rangle }{T}
- \frac{\langle \Delta E_1^{\kappa\chi}\Delta E_2^{\alpha\beta\nu\eta}\rangle }{T}\ .
\end{split}\end{equation}
The third order elastic constants are thus given by
\begin{equation}\begin{split}
\tilde{C}_3^{\alpha\beta\nu\eta\kappa\chi} & = \frac{1}{V}\left[
\langle E_3^{\alpha\beta\nu\eta\kappa\chi}\rangle
+\frac{\langle \Delta E_1^{\alpha\beta}\Delta E_1^{\nu\eta}\Delta E_1^{\kappa\chi} \rangle}{T^2} \right. \\
&\quad\quad\quad - \left.\frac{\langle \Delta E_1^{\alpha\beta}\Delta E_2^{\nu\eta\kappa\chi}\rangle }{T}
- \frac{\langle \Delta E_1^{\nu\eta}\Delta E_2^{\alpha\beta\kappa\chi}\rangle }{T}
- \frac{\langle \Delta E_1^{\kappa\chi}\Delta E_2^{\alpha\beta\nu\eta}\rangle }{T}\right]\ .
\label{3rdorder}
\end{split}\end{equation}
\end{widetext}

%%%%%%%%%%%%%%%%%%%%%%%%%%%%%%%%%%%%%%%%%%%%%%%%%%%%%%%%%%%%%%%%%%%%%%%%%%%%%%%%%
\section{$T \to 0$ limit of thermal elastic coefficients}
\label{limit}
We first work out a general expression for the low temperature expansion of the mean of any quantity $A(\xv)$ which depends only on coordinates; we start with the definition
\begin{equation}\label{foo7}
\langle A \rangle = \frac{1}{{\cal Z}_0}\int A e^{-E/T} \,d\B pd\B q =
\frac{1}{{\cal Z}_c}\int A e^{-U/T} \,d\B q\ ,
\end{equation}
where ${\cal Z}_c = \int e^{-\frac{U}{T}} d\B q$.
With the notation $\delta x^i_\nu \equiv x^i_\nu - \tilde{x}^i_\nu$,
we expand $A$ around some local minimum $\tilde{\xv}$,
up to fourth order in coordinates
\begin{eqnarray}\label{foo11}
A  & \simeq &  A(\tilde{\xv}) +
\left.\frac{\partial A}{\partial x^i_\nu}\right|_{\tilde{\xv}}
\delta x^i_\nu+\frac{1}{2}\left.\frac{\partial^2 A}
{\partial x_\eta^j\partial x_\nu^i}
\right|_{\tilde{\xv}}\delta x^i_\nu \delta x^j_\eta \nonumber \\ && +
\frac{1}{6}\!\!\left.\frac{\partial^3 A}{
\partial x_\zeta^\ell\partial x_\eta^j\partial x_\nu^i}
\right|_{\tilde{\xv}}\!\!\!\! \delta x^i_\nu \delta x^j_\eta \delta x^\ell_\zeta \\ && +
\frac{1}{24}\left.\frac{\partial^4 A}
{\partial x_\theta^m\partial x_\zeta^\ell
\partial x_\eta^j\partial x_\nu^i}
\right|_{\tilde{\xv}}\!\!\!\! \delta x^i_\nu \delta x^j_\eta
\delta x^\ell_\zeta \delta x^m_\theta + \ldots\ \nonumber \ .
\end{eqnarray}
We now expand the potential energy up to fifth order in coordinates
around the local minimum $\tilde{\xv}$, with
$\calBold{T},\calBold{M},\calBold{Q}$ denoting the third, fourth and fifth
order derivatives of the potential energy with respect to coordinates,
respectively, evaluated at $\tilde{\xv}$:
\begin{equation}\begin{split}
U \simeq U(\tilde{\xv}) & + \sFrac{1}{2}{\cal H}^{ij}_{\eta\nu}
\delta x^i_\nu \delta x^j_\eta + \sFrac{1}{6}
{\cal T}^{ij\ell}_{\nu\eta\zeta}\delta x^i_\nu \delta x^j_\eta \delta x^\ell_\zeta  \\ &
+ \sFrac{1}{24}{\cal M}^{ij\ell m}_{\nu\eta\zeta\theta}\delta x^i_\nu \delta x^j_\eta
\delta x^\ell_\zeta \delta x^m_\theta \\ & +
\sFrac{1}{120}{\cal Q}^{ij\ell m p}_{\nu\eta\zeta\theta\tau}
\delta x^i_\nu \delta x^j_\eta\delta x^\ell_\zeta \delta x^m_\theta \delta x^p_\tau \ ,
\end{split}\end{equation}
where the first order term vanishes since
$\left.\frac{\partial U}{\partial x^i_\nu}\right|_{\tilde{\xv}} = 0$ for
every $x^i_\nu$. At low temperatures the virial theorem insures that the second order term in the
potential energy expansion is proportional to $T$, hence the third, fourth and fifth order terms are
of higher order in temperature; thus, in the athermal limit, we can
expand the corresponding parts of the Boltzmann factor:
\begin{widetext}
\begin{equation}\begin{split}
e^{-\frac{U}{T}} \simeq
e^{-\frac{U(\tilde{\xv}) + \sFrac{1}{2}{\cal H}_{\nu\eta}^{ij}
\delta x^i_\nu \delta x^j_\eta}{T}}\!
&\left[ 1 - \frac{{\cal T}^{ij\ell}_{\nu\eta\zeta}
\delta x^i_\nu \delta x^j_\eta \delta x^\ell_\zeta}{6T}
 - \frac{{\cal M}^{ij\ell m}_{\nu\eta\zeta\theta}
\delta x^i_\nu \delta x^j_\eta
\delta x^\ell_\zeta \delta x^m_\theta}{24T}\right. \\ &\ -
\frac{{\cal Q}^{ij\ell m p}_{\nu\eta\zeta\theta\tau}
\delta x^i_\nu \delta x^j_\eta\delta x^\ell_\zeta
\delta x^m_\theta \delta x^p_\tau}{120T} \left.
+\frac{{\cal T}_{\nu\eta\zeta}^{ij\ell}{\cal T}^{mpq}_{\theta\tau\rho}
\delta x^i_\nu \delta x^j_\eta\delta x^\ell_\zeta
\delta x^m_\theta \delta x^p_\tau\delta x^q_\rho}{72T^2} + \ldots\right] .
\end{split}\end{equation}
We have omitted the other sixth order term since it does not eventually
contribute to the temperature expansion of $\langle A\rangle$.
Inserting the above expansion and (\ref{foo11}) into (\ref{foo7}), we carry out the
Gaussian integrals to obtain the low temperature approximation
of the equilibrium mean of $A$ up to terms of ${\cal O}(T^3)$:
\begin{equation}\label{generalAthermalLimit}
\begin{split}
&\langle A\rangle \simeq  \ A(\tilde{\xv}) + \frac{T}{2}\left[
\calBold{H}^{-1} \cdot \calBold{A}_{\xv\xv}
- \calBold{H}^{-1} \cdot\calBold{T} \cdot \calBold{H}^{-1}
\cdot \calBold{A}_{\xv}\right]
\\&  + \frac{T^2}{48}\left[ 6\calBold{H}^{-1} \cdot
\calBold{A}_{\xv\xv\xv\xv} \cdot \calBold{H}^{-1}-
20\calBold{H}^{-1} \cdot \calBold{T} \cdot (\calBold{H}^{-1}  \calBold{H}^{-1})
\cdot \calBold{A}_{\xv\xv\xv}
-3\calBold{H}^{-1} \cdot
\calBold{T}\cdot \calBold{H}^{-1}\cdot \calBold{A}_{\xv}\calBold{H}^{-1}\cdot
\calBold{M}\cdot\calBold{H}^{-1}
\right. \\&\
-6(\calBold{H}^{-1}\calBold{H}^{-1})\cdot
\calBold{Q} \cdot\calBold{H}^{-1}\cdot\calBold{A}_\xv
\left.  -12\calBold{H}^{-1}\cdot \calBold{A}_{\xv\xv}\calBold{H}^{-1} \cdot
\calBold{M}\cdot \calBold{H}^{-1}
-40(\calBold{H}^{-1}\cdot\calBold{T})\cdot (\calBold{H}^{-1}\cdot \calBold{T}
\cdot\calBold{H}^{-1})(\calBold{H}^{-1}\cdot\calBold{A}_{\xv\xv})
\right]\ ,
\end{split}
\end{equation}
where ${\calBold {A}}_{\B x}$, $\calBold {A}_{\B {xx}}$ etc. denote the tensors of first, second etc. derivatives of $A$
with respect to $\B x$.
Using this expression, we derive relations for fluctuations. Assume that
also $B$ and $C$ are functions of coordinates $\xv$; with the definition
$\Delta A = A - \langle A\rangle$, we have
\begin{equation}\label{secondOrderFlucs}
\frac{\langle \Delta A \Delta B \rangle}{T} \simeq
\sFrac{1}{2}\left[\calBold{A}_\xv \cdot \calBold{H}^{-1}\cdot\calBold{B}_\xv
+\calBold{B}_\xv\cdot\calBold{H}^{-1}\cdot\calBold{A}_\xv\right] + {\cal O}(T)\ ,
\end{equation}
and
\begin{equation}
\label{thirdOrderFlucs}
\begin{split}
\frac{\langle \Delta A \Delta B \Delta C \rangle}{T^2} \simeq &\quad
\calBold{H}^{-1} \cdot \calBold{A}_{\xv}\cdot \calBold{C}_{\xv\xv}\cdot \calBold{H}^{-1}
\cdot \calBold{B}_{\xv}
+\calBold{H}^{-1}\cdot \calBold{B}_{\xv}\cdot \calBold{A}_{\xv\xv}\cdot\calBold{H}^{-1}
\cdot \calBold{C}_{\xv} \\ &\
+\calBold{H}^{-1} \cdot \calBold{C}_{\xv}\cdot \calBold{B}_{\xv\xv}\cdot \calBold{H}^{-1}
\cdot \calBold{A}_{\xv}
- (\calBold{H}^{-1}\cdot \calBold{A}_{\xv}) (\calBold{H}^{-1}\cdot \calBold{B}_{\xv})
(\calBold{H}^{-1}\cdot \calBold{C}_{\xv})\cdot\calBold{T} + {\cal O}(T)\ .
\end{split}\end{equation}
In this section we will make use of the following
definitions, (see also (\ref{E_terms})):
\begin{equation}\label{definitions}\begin{split}
U_0 &\ \equiv U(\B \epsilon = 0) \ , \
U_1^{\alpha\beta} \equiv \left.\frac{\partial U}
{\partial \epsilon_{\alpha\beta}}\right|_{\B \epsilon = 0} \ , \
U_2^{\alpha\beta\nu\eta} \equiv \left.\frac{\partial^2 U}
{\partial \epsilon_{\nu\eta}\partial \epsilon_{\alpha\beta}}\right|_{\B \epsilon = 0} \ , \
U_3^{\alpha\beta\nu\eta\kappa\chi} \equiv \left.\frac{\partial^3 U}
{\partial \epsilon_{\kappa\chi}\partial \epsilon_{\nu\eta}
\partial \epsilon_{\alpha\beta}}\right|_{\B \epsilon = 0} \ , \\
K_0 &\ \equiv K(\B \epsilon = 0) \ , \
K_1^{\alpha\beta} \equiv \left.\frac{\partial K}
{\partial \epsilon_{\alpha\beta}}\right|_{\B \epsilon = 0} \ , \
K_2^{\alpha\beta\nu\eta} \equiv \left.\frac{\partial^2 K}
{\partial \epsilon_{\nu\eta}\partial \epsilon_{\alpha\beta}}\right|_{\B \epsilon = 0} \ , \
K_3^{\alpha\beta\nu\eta\kappa\chi} \equiv \left.\frac{\partial^3 K}
{\partial \epsilon_{\kappa\chi}\partial \epsilon_{\nu\eta}\partial \epsilon_{\alpha\beta}}
\right|_{\B \epsilon = 0} \\
E_0 &\ \equiv U_0 +  K_0 \ , \
E_1^{\alpha\beta} \equiv U_1^{\alpha\beta} + K_1^{\alpha\beta} \ , \
E_2^{\alpha\beta\nu\eta} \equiv U_2^{\alpha\beta\nu\eta} +
K_2^{\alpha\beta\nu\eta}\ , \
E_3^{\alpha\beta\nu\eta\kappa\chi} \equiv U_3^{\alpha\beta\nu\eta\kappa\chi}
+ K_3^{\alpha\beta\nu\eta\kappa\chi}\ .
\end{split}\end{equation}
\end{widetext}

\subsection{First order elastic constants -- athermal limit}
The first order free energy derivative with respect to strain is given by
(see Eqs.~(\ref{E_terms}),(\ref{firstOrderDerivative}))
\begin{equation}
\frac{\partial {\cal F}}{\partial \epsilon_{\alpha\beta}}
 =  \langle E_1^{\alpha\beta} \rangle = \langle U_1^{\alpha\beta} +
K_1^{\alpha\beta}\rangle = \langle U_1^{\alpha\beta}\rangle + \langle K_1^{\alpha\beta}\rangle.
\end{equation}
In the athermal limit the kinetic term (\ref{firstDerivativeKinetic}) vanishes and
following relation (\ref{generalAthermalLimit}) we are left with
\begin{equation}\label{foo6}
\lim_{T\to 0}\langle E_1^{\alpha\beta} \rangle = \lim_{T\to 0}\langle U_1^{\alpha\beta} \rangle
= \left.\frac{\partial U}
{\partial \epsilon_{\alpha\beta}}\right|_{\tilde{\xv}} \ .
\end{equation}
From here the athermal limit of the first order elastic constants is
\begin{equation}
\lim_{T\to 0}\tilde{C}_1^{\alpha\beta} = \frac{1}{V}\left.\frac{\partial U}
{\partial \epsilon_{\alpha\beta}}\right|_{\tilde{\xv}}\ ,
\end{equation}
in agreement with (\ref{firstOrderAthermal}).

\subsection{Second order elastic constants -- athermal limit}
The second order free energy derivative with respect to strain is given by
(see Eqs.~(\ref{E_terms}),(\ref{secondOrderDerivative}))
\begin{equation}\label{moo5}
\frac{\partial ^2{\cal F}}{ \partial \epsilon_{\nu\eta}
\partial\epsilon_{\alpha\beta}} = \langle E_2^{\alpha\beta\nu\eta}\rangle
-\sFrac{1}{T}\langle \Delta E_1^{\alpha\beta}\Delta E_1^{\nu\eta}\rangle\ .
\end{equation}
The first term is
$\langle E_2^{\alpha\beta\nu\eta}\rangle =
\langle U_2^{\alpha\beta\nu\eta}\rangle +
\langle K_2^{\alpha\beta\nu\eta}\rangle$; in the athermal
limit the kinetic term (\ref{secondDerivativeKinetic}) vanishes, and
following (\ref{generalAthermalLimit}) we are left with
\begin{equation}
\lim_{T\to 0}\left< \frac{\partial ^2U}{ \partial \epsilon_{\nu\eta}
\partial\epsilon_{\alpha\beta}}\right> =
\left.\frac{\partial ^2U}{ \partial \epsilon_{\nu\eta}
\partial\epsilon_{\alpha\beta}}\right|_{\tilde{\xv}}\ .
\end{equation}
The second term in (\ref{moo5}) is
\[\begin{split}
\frac{\langle \Delta E_1^{\alpha\beta}\Delta E_1^{\nu\eta}\rangle}{T}  = &\
\frac{\langle (\Delta U_1^{\alpha\beta} + \Delta K_1^{\alpha\beta})
( \Delta U_1^{\nu\eta} + \Delta K_1^{\nu\eta}) \rangle}{T} \\
= &\ \frac{1}{T}\left[
\langle \Delta U_1^{\alpha\beta}\Delta U_1^{\nu\eta}\rangle +
\langle \Delta U_1^{\alpha\beta}\Delta K_1^{\nu\eta}\rangle  \right.\\ & \  +
\langle \Delta K_1^{\alpha\beta}\Delta U_1^{\nu\eta}\rangle +
\left.\langle \Delta K_1^{\alpha\beta}\Delta K_1^{\nu\eta}\rangle \right]\ .
\end{split}\]
Notice that
\begin{eqnarray}
\langle \Delta U_1^{\alpha\beta}\Delta K_1^{\nu\eta}\rangle & = &
\frac{1}{{\cal Z}}\int e^{-\frac{U + K}{T}}
\Delta U_1^{\alpha\beta}\Delta K_1^{\nu\eta}\ d\B pd\B q \nonumber \\
& = &\frac{1}{{\cal Z}}\int e^{-\frac{U}{T}}
\Delta U_1^{\alpha\beta}\ d\B q\int e^{-\frac{K}{T}}\Delta K_1^{\nu\eta}\ d\B p
\nonumber \\ & = &
\langle \Delta U_1^{\alpha\beta}\rangle\langle\Delta K_1^{\nu\eta}\rangle = 0\ , \nonumber
\end{eqnarray}
so the second term in (\ref{moo5}) reduces to
\[
\frac{\langle \Delta E_1^{\alpha\beta}\Delta E_1^{\nu\eta}\rangle}{T}  =
\frac{\langle \Delta U_1^{\alpha\beta}\Delta U_1^{\nu\eta}\rangle +
\langle \Delta K_1^{\alpha\beta}\Delta K_1^{\nu\eta}\rangle}{T}.
\]
The second term of the RHS of the above relation is proportional to $T^2$,
so we are left with
\[
\lim_{T\to 0}
\sFrac{1}{T}\langle \Delta E_1^{\alpha\beta}\Delta E_1^{\nu\eta}\rangle
= \lim_{T\to 0}\sFrac{1}{T}
\langle \Delta U_1^{\alpha\beta}
\Delta U_1^{\nu\eta}\rangle\ .
\]
Using the relation (\ref{secondOrderFlucs}) and definition (\ref{nonAffVel}),
we obtain
\begin{equation}\begin{split}
\lim_{T\to 0}\sFrac{1}{T}
\langle \Delta U_1^{\alpha\beta}
\Delta U_1^{\nu\eta}\rangle = &\
\sFrac{1}{2}\left[\Xi^i_{\kappa\alpha\beta}({\cal H}^{-1})^{ij}_{\kappa\chi}
\Xi^j_{\chi\nu\eta} \right. \\ &\ \   +
\left. \Xi^i_{\kappa\nu\eta}({\cal H}^{-1})^{ij}_{\kappa\chi}
\Xi^j_{\chi\alpha\beta}\right] \\  = &\  -{\cal V}^i_{\chi\alpha\beta}
\Xi^i_{\chi\nu\eta}\ .
\end{split}\end{equation}
Finally
\begin{equation}
\lim_{T\to 0}\tilde{C}_2^{\alpha\beta\nu\eta} = \frac{1}{V}\left[
\left.\frac{\partial ^2U}{ \partial \epsilon_{\nu\eta}
\partial\epsilon_{\alpha\beta}}\right|_{\tilde{\xv}} +
{\cal V}^i_{\chi\alpha\beta}\Xi^i_{\chi\nu\eta}\right]\ .
\end{equation}
in agreement with (\ref{secondOrderElasticConstants}).

\subsection{Third order elastic constants -- athermal limit}
The third order free energy derivative with respect to strain is given by
(see Eqs.~(\ref{E_terms}),(\ref{thirdOrderDerivative}))
\begin{widetext}
\begin{equation}\label{foo8}\begin{split}
\frac{\partial ^3{\cal F}}{\partial \epsilon_{\kappa\chi} \partial \epsilon_{\nu\eta}
\partial\epsilon_{\alpha\beta}} & = \langle E_3^{\alpha\beta\nu\eta\kappa\chi}\rangle
+\frac{\langle \Delta E_1^{\alpha\beta}\Delta E_1^{\nu\eta}
\Delta E_1^{\kappa\chi} \rangle}{T^2}  \\
&\quad\quad\quad - \frac{\langle \Delta E_1^{\alpha\beta}\Delta E_2^{\nu\eta\kappa\chi}\rangle }{T}
- \frac{\langle \Delta E_1^{\nu\eta}\Delta E_2^{\alpha\beta\kappa\chi}\rangle }{T}
- \frac{\langle \Delta E_1^{\kappa\chi}\Delta E_2^{\alpha\beta\nu\eta}\rangle }{T}\ .
\end{split}\end{equation}
The first term on the RHS of (\ref{foo8}) is
$\langle E_3^{\alpha\beta\nu\eta\kappa\chi}\rangle =
\langle U_3^{\alpha\beta\nu\eta\kappa\chi}\rangle +
\langle K_3^{\alpha\beta\nu\eta\kappa\chi}\rangle$; in the athermal
limit the kinetic term (\ref{thirdDerivativeKinetic}) vanishes, and
following (\ref{generalAthermalLimit}) we are left with
\begin{equation}\label{moo10}
\lim_{T\to 0}\left< \frac{\partial ^3U}{ \partial \epsilon_{\kappa\chi}
\partial \epsilon_{\nu\eta}\partial\epsilon_{\alpha\beta}}\right> =
\left.\frac{\partial ^3U}{\partial \epsilon_{\kappa\chi} \partial \epsilon_{\nu\eta}
\partial\epsilon_{\alpha\beta}}\right|_{\tilde{\xv}}\ .
\end{equation}
The second term in (\ref{foo8}) is
\begin{equation}
\frac{\langle \Delta E_1^{\alpha\beta}\Delta E_1^{\nu\eta}
\Delta E_1^{\kappa\chi} \rangle}{T^2} =
\frac{\langle  \Delta U_1^{\alpha\beta}\Delta U_1^{\nu\eta}\Delta U_1^{\kappa\chi}
\rangle + \langle\Delta K_1^{\alpha\beta}
\Delta K_1^{\nu\eta}\Delta K_1^{\kappa\chi}\rangle}{T^2}\ ,
\end{equation}
since similarly to the second order athermal limit case, it is easy to
verify that all mixed terms envolving
products of kinetic energy and potential energy derivative cancel.
The triple product $\langle\Delta K_1^{\alpha\beta}
\Delta K_1^{\nu\eta}\Delta K_1^{\kappa\chi}\rangle \sim T^3$, so we are left with
\[
\lim_{T\to 0}\frac{\langle \Delta E_1^{\alpha\beta}\Delta E_1^{\nu\eta}
\Delta E_1^{\kappa\chi} \rangle}{T^2} = \lim_{T\to 0}
\frac{\langle  \Delta U_1^{\alpha\beta}\Delta U_1^{\nu\eta}\Delta U_1^{\kappa\chi}
\rangle}{T^2}\ .
\]
Using relation (\ref{thirdOrderFlucs}) and definition (\ref{nonAffVel}),
this is
\begin{equation}\label{moo11}
\lim_{T\to 0}
\frac{\langle \Delta U_1^{\alpha\beta}\Delta U_1^{\nu\eta}
\Delta U_1^{\kappa\chi}\rangle}{T^2} =
{\cal V}^i_{\theta\alpha\beta}{\cal V}^j_{\zeta\nu\eta}
{\cal V}^\ell_{\rho\kappa\chi}{\cal T}^{ij\ell}_{\theta\zeta\rho}
+ {\cal V}^i_{\theta\alpha\beta}{\cal V}^j_{\zeta\nu\eta}
\frac{\partial \Xi^j_{\zeta\kappa\chi}}{\partial X^i_\theta}
+ {\cal V}^i_{\theta\nu\eta}{\cal V}^j_{\zeta\kappa\chi}
\frac{\partial \Xi^j_{\zeta\alpha\beta}}{\partial X^i_\theta}
+ {\cal V}^i_{\theta\kappa\chi}{\cal V}^j_{\zeta\nu\eta}
\frac{\partial \Xi^j_{\zeta\alpha\beta}}{\partial X^i_\theta}\ .
\end{equation}

Finally, we use relation (\ref{secondOrderFlucs}) for the remaining terms:
\begin{equation}\label{moo12}
\lim_{T\to 0}\frac{\langle \Delta E_1^{\alpha\beta}
\Delta E_2^{\nu\eta\kappa\chi}\rangle }{T} =
\lim_{T\to 0}\frac{\langle \Delta U_1^{\alpha\beta}
\Delta U_2^{\nu\eta\kappa\chi}\rangle }{T} =
-{\cal V}^i_{\zeta\alpha\beta}\frac{\partial \Xi^i_{\zeta\nu\eta}}
{\partial \epsilon_{\kappa\chi}}\ .
\end{equation}
Combining results (\ref{moo10}),(\ref{moo11}),(\ref{moo12}), we arrive at
the final result

\begin{equation}\begin{split}
\lim_{T\to 0}\tilde{C}_3^{\alpha\beta\nu\eta\kappa\chi}
= \ \frac{1}{V}&\left[\frac{\partial ^3 U}{\partial \epsilon_{\kappa\chi}\partial \epsilon_{\nu\eta}
\partial \epsilon_{\alpha\beta}} +
{\cal V}^i_{\theta\alpha\beta}{\cal V}^j_{\zeta\nu\eta}
{\cal V}^\ell_{\rho\kappa\chi}\frac{\partial^3 U}
{\partial X^i_\theta \partial X^j_\zeta \partial X^\ell_\rho} \right. \\
&\  + {\cal V}^i_{\theta\alpha\beta}{\cal V}^j_{\zeta\nu\eta}
\frac{\partial \Xi^j_{\zeta\kappa\chi}}{\partial X^i_\theta}
+ {\cal V}^i_{\theta\nu\eta}{\cal V}^j_{\zeta\kappa\chi}
\frac{\partial \Xi^j_{\zeta\alpha\beta}}{\partial X^i_\theta}
+ {\cal V}^i_{\theta\kappa\chi}{\cal V}^j_{\zeta\nu\eta}
\frac{\partial \Xi^j_{\zeta\alpha\beta}}{\partial X^i_\theta}
\\ &\  +{\cal V}^i_{\zeta\alpha\beta}\frac{\partial \Xi^i_{\zeta\nu\eta}}
{\partial \epsilon_{\kappa\chi}} +
{\cal V}^i_{\zeta\nu\eta}\frac{\partial \Xi^i_{\zeta\kappa\chi}}
{\partial \epsilon_{\alpha\beta}} +
\left. {\cal V}^i_{\zeta\kappa\chi}\frac{\partial \Xi^i_{\zeta\alpha\beta}}
{\partial \epsilon_{\nu\eta}} \right]\ .
\end{split}\end{equation}
in agreement with (\ref{thirdOrderAthermalDerivative}).
\end{widetext}

%%%%%%%%%%%%%%%%%%%%%%%%%%%%%%%%%%%%%%%%%%%%%%%%%%%%%%%%%%%%%%%%%%
\section{Examples of applications}
\label{applications}

To justify and motivate the calculation of the nonlinear elastic constants we present now two examples of important issues
regarding elasto-plastic behavior in amorphous solids that cannot be discussed without invoking these nonlinear constants.
The first is plasticity-induced anisotropy and the second is the predictions of plastic failure.
%%%%%%%%%%%%%%%%%%%%%%%%%%%%
\subsection{Plasticity-induced anisotropy}

\begin{figure}
\centering
\includegraphics[scale = 0.45]{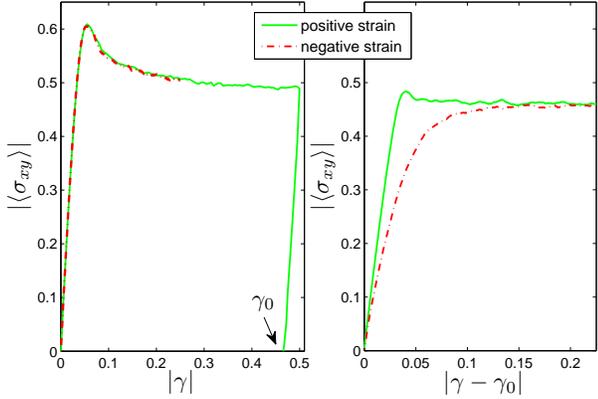}
\caption{Color online: Stress-strain curves. Left panel: starting the experiment
from a freshly prepared sample results in a symmetric trajectory for $\gamma\to -\gamma$.
Right panel: starting the experiment from the zero-stress state with $\gamma=\gamma_0$
results in an asymmetric trajectory, see text for details. Data was averaged over
500 independent stress-strain curves at $T=0.01$ where temperature is measured
in units of $\varepsilon/k_B$, see \cite{10KLP} for details}
\label{stress-strain}
\end{figure}

A freshly produced amorphous solid is isotropic, and as such presents a symmetric stress vs.~strain curve for positive or
negative strain. This is not the case for the same amorphous solid after it had been already strained such that its stress exceeded its yield-stress where plastic deformations become numerous.
This is demonstrated in Fig.~\ref{stress-strain}.
%%%%%%%%%%%%%%%%%%%%%%%%%%%%%%%%%%%%%%%%%%%%%%%%%%%%%%
%%%%%%%%%%%%%%%%%%%%%%%%%%%%%%%%%%%%%%%%%%%%%%%%%
A typical averaged stress-strain curve for a model
amorphous solid (see Ref.~\cite{10KLP} for numerical details) starting from an ensemble of freshly prepared homogenous states is shown in the left panel, with a symmetric trajectory for positive or negative shear strain. Once in the steady flow state, each system in the ensemble is brought back to a zero-stress state, which serves as the starting point for a second experiment in which a positive and negative strain is put on the system as shown
in the right panel of Fig.~\ref{stress-strain}. Even though the initial ensemble is prepared to have zero mean stress, the
average trajectory is now asymmetric. This phenomenon, sometime referred to as the Bauschinger effect \cite{90BCH},
shows that the starting point $\gamma_0$ for the second experiment (referred below as the Bauschinger point) retains a memory of the loading history in some form of anisotropy.

\begin{figure}
 \centering
\includegraphics[scale = 0.57]{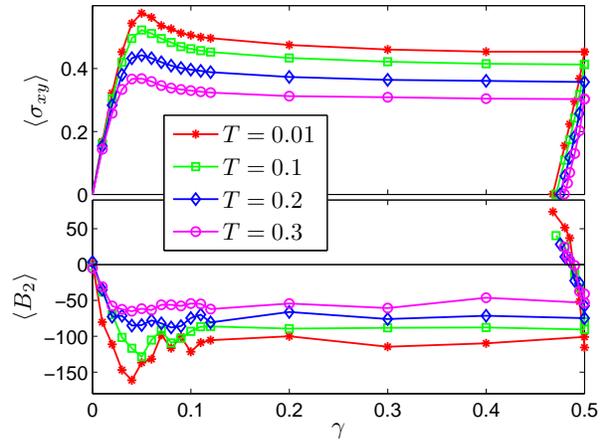}
\caption{Color Online: Upper panel: Trajectories of stress vs. strain for four different temperature at the same strain rate $\gamma=10^{-4}$. Lower panel: the corresponding values of $B_2$ as a function of strain. Data was averaged over 1000 independent stress-strain curves at each temperature. Note that $B_2$ is negative even when the averaged stress-strain curve has a positive curvature, see text for discussion.}
 \label{B2vsgamma}
 \end{figure}

To shed light on the anisotropy of the Bauschinger point we choose to measure the sum
\begin{equation}
B_2(\gamma^*) \equiv \!\lim_{T\to 0}\! \left[\tilde{C}_2^{xyyy} + \tilde{C}_3^{xyxyxy}\right] \!= \!\lim_{T\to 0}\left. \frac{d^2 \sigma_{xy}}{d \gamma^2}
\right|_{\gamma = \gamma^*},
\label{defB2}
\end{equation}
which can be determined using the results Eqs. (\ref{secondOrderElasticConstants}) and (\ref{3rdorder}). In particular
we note that $B_2$ is identically zero in an isotropic ensemble. It gains a nonzero value when plastic events take place
and begin to build anisotropy. In Fig.~\ref{B2vsgamma} we present results of numerical simulations in a typical
model of an amorphous solid (for details cf.~\cite{10KLP}), and present the measured value of $B_2$ along the trajectory
%%%%%%%%%%%%%%%%%%%%%%%%%%%%%%%%%%%%%%%%%%%%%
%%%%%%%%%%%%%%%%%%%%%%%%%%%%%%%
shown in Fig.~\ref{stress-strain}. In addition to the very low temperature trajectory in Fig.~\ref{stress-strain} we show also measurements of $B_2$ for simulations performed at other three different temperatures. For all these trajectories
$B_2$ was measured by quenching to zero temperature configurations sampled along the thermal trajectory. The magnitude of the Bauschinger effect
goes down when temperature goes up, and this is in good agreement with the value of $B_2$ at the Bauschinger point which
also decreases when temperature increases. We thus see that the values of the nonlinear elastic constants can serve as
natural measures for the degree of anisotropy that is built up in an amorphous solid due to plastic deformations.
Further discussion of this measure can be found in \cite{10KLP}. Note that in \cite{10KLP} the value of $B_2$ was obtained
directly from stress vs. strain curves, computing the derivatives numerically. With the results obtained in this paper
we can compute $B_2$ or any other elastic constant directly from the particle interactions. This will open up in the future
a possibility to define local values of the elastic constants, providing maps of $B_2$ or other nonlinear elastic constants.

%%%%%%%%%%%%%%%%%%%%%%%%%%%%%%%%%%%%%%%%%%%%%
\subsection{Predicting plasticity}

Imagine an amorphous solid under a given state of strain. Can one predict how much additional strain
is needed to reach plastic failure? Recently we have shown, cf.~\cite{10KLLP}, that an accurate predictor of plastic
failure in an amorphous solid can be constructed
with the help of the higher order derivatives of the potential function. Our findings not only offer a predictive
tool for the onset of failure, but also point out the
importance of nonlinearities, and in particular those that couple
nonlinearly ``softening'' regions with strain at larger
scales. It is indeed this non-linear interaction which produces
a reinforcing mechanism leading ultimately to a
catastrophic event in the form of plastic yielding.

To fix ideas, imagine a simple shear deformation applied to a given
piece of amorphous solid (for simplicity in 2D, with immediate extensions to 3D).
A small strain increment $\delta\gamma$ corresponds to a change of the $i$'th particle
positions $\B x_i\to \B x'_i$ as:
\begin{eqnarray}
x'_i & =& x_i + \delta\gamma y_i \ , \nonumber \\
y'_i & =& y_i \nonumber \ ,
\end{eqnarray}
In athermal quasi-static conditions ($T\to 0, \quad \dot\gamma\to 0$), the system lives
in local minima, and follows strain-induced changes of the potential energy surface.
Therefore, the particles do not follow homogeneously the macroscopic strain, and their
positions change as $\B x_i\to \B x'_i+\Xv_i$, where $\Xv_i$ denotes non-affine displacements.
Around some stable reference state at $\gamma=\gamma^*$,
the field $\Xv_i$, the system energy, and internal stress $\sigma_{xy}$ are smooth functions of $\gamma$. We can thus write:
\begin{equation}
\sigma_{xy} (\gamma) \!= \!\sum_{n=0}^\infty \frac{B_n}{n!} (\gamma-\gamma^*)^n\ ,
~~ B_n =\lim_{T\to 0}\left. \frac{d ^n \sigma_{xy}}{d \gamma^n}
\right|_{\gamma = \gamma^*}. \label{series}
\end{equation}
As the strain increases, the system must eventually lose mechanical stability;
the ``elastic branch'' on the stress curve ends in a discontinuity as the system
fails via a first subsequent ``plastic event''.  It is precisely at this instability,
say at $\gamma=\gamma_P$, that the function $\sigma_{xy}(\gamma)$ loses
its analyticity. Accordingly we recognize that the radius of convergence of the
series (\ref{series}) is precisely $|\gamma_P-\gamma_0|$, where $\gamma_P$
can be larger or smaller than $\gamma^*$.

Mechanical instabilities are associated with
the vanishing of an eigenvalue of the
Hessian, which we will denote as $\lambda_P$.
Here, as opposed to the analysis of \cite{04ML,10KLLP},
we do not make any assumptions about the functional form
of $\lambda_P(\gamma)$. The vanishing of $\lambda_P$ implies that
any terms of the elastic constants that are composed of the inverse of
the Hessian $\calBold{H}^{-1}$ will diverge at $\lambda_P$; hence,
the strongest diverging terms are those which are composed of the largest number
of $\calBold{H}^{-1}$'s. Accordingly, higher order elastic
constants will consist of stronger divergences,
as can be understood, for example, by comparing Eq.~(\ref{secondOrderElasticConstants})
with Eq.~(\ref{thirdOrderAthermalDerivative}). Near $\gamma_P$ the most diverging terms will dominate
over the rest, so we will consider in the following analysis the most diverging terms
of each order of the athermal elastic constants.

We begin with $B_1 = C_2^{xyxy}$, which is given by Eq.~(\ref{secondOrderElasticConstants});
following the discussion above, the most (and only, for $\calBold{C}_2$) diverging
term is ${\cal V}^i_{\theta xy}\Xi^i_{\theta xy}/V$ (recall that
$\calBold{V}$ consists of $\calBold{H}^{-1}$, see Eqs.~(\ref{nonAffVel}),(\ref{ModeExp})).
Close to $\gamma_P$, the diverging term will dominate, so we can write
\begin{equation}
B_1 \sim \frac{1}{\lambda_P}\ .
\end{equation}
We continue with $B_2$; the most
diverging term in $B_2$ is the most diverging term in $C_3^{xyxyxy}$; from
Eq.~(\ref{thirdOrderAthermalDerivative}), this is
\begin{equation}
B_2 \sim {\cal V}^i_{\theta xy}{\cal V}^j_{\zeta xy}
{\cal V}^\ell_{\rho xy}{\cal T}^{ij\ell}_{\theta\zeta\rho} \sim \frac{1}{\lambda_P^3}\ .
\end{equation}
Since $B_2 \equiv \frac{d B_1}{d\gamma}$, we obtain a differential
equation for $\lambda_P(\gamma)$:
\[
\frac{1}{\lambda_P^3} \sim \frac{d}{d\gamma}\left(\frac{1}{\lambda_P}\right)
= -\frac{1}{\lambda_P^2}\frac{d\lambda_P}{d\gamma}\ .
\]
We re-write this as
\begin{equation}
\frac{d \lambda_P}{d\gamma} \sim -\frac{1}{\lambda_P}\ ,
\end{equation}
for which, together with the boundary condition $\lambda_P|_{\gamma_P} = 0$,
the solution is
\begin{equation}
\lambda_P \sim \sqrt{\gamma_P - \gamma}\ ,
\end{equation}
in agreement with \cite{04ML,10KLLP}.

With this result, we are able to derive expressions for the
diverging terms of $B_3$ and $B_4$, as a function of $\gamma_P$;
starting from $B_1 = \frac{a}{\sqrt{\gamma_P - \gamma}}$
we obtain the relations
\begin{equation}
B_3 \simeq \frac{3a}{4(\gamma_P - \gamma)^{\frac{5}{2}}}\quad \mbox{and} \quad
B_4 \simeq \frac{15a}{8(\gamma_P - \gamma)^{\frac{7}{2}}}\ .
\end{equation}
Solving for $\gamma_P$, we obtain the prediction
\begin{equation}
\gamma_P = \gamma + \frac{5 B_3}{2B_4}\ ,
\end{equation}
where the most diverging terms of $B_3$ and $B_4$ should be considered.
Notice that one could, in principle derive expressions for $\gamma_P$
involving lower order elastic constants; see \cite{10KLLP} for discussion.

We finally derive expressions for the most diverging terms of $B_3$ and
$B_4$; starting from the most diverging term in $\calBold{C}_3$, we take
another constrained derivative:
\begin{eqnarray}
\left.\frac{\partial}{\partial \calBold{\epsilon} }\left(
[\calBold{V}\calBold{V}\calBold{V}]\cdot\calBold{T}\right)\right|_{\fv}
& = & -\left[\calBold{T}\cdot\calBold{V}
\calBold{V}\right]\cdot\calBold{H}^{-1}\cdot
\left[\calBold{T}\cdot\calBold{V}\calBold{V}\right] \nonumber \\
&& \ \  + \mbox{2 symmetric terms}\ , \label{zoo3}
\end{eqnarray}
where here and in the following, the contractions
are \emph{only} over indices and components of particle coordinates,
and we only consider here the most diverging term of
$\left.\frac{\partial \calBold{V}}{\partial \calBold{\epsilon} }\right|_{\fv}$, see
Eq.~(\ref{nonAffineAcc}).

Taking another constrained derivative of (\ref{zoo3})
requires an expression for
$\left.\frac{\partial ({\cal H}^{-1})^{ij}_{\alpha\beta}}
{\partial \epsilon_{\nu\eta} }\right|_{\fv}$, which can be obtained by
applying the rule (\ref{constDer2}) on $\calBold{H}^{-1}\cdot\calBold{H}$:
\begin{eqnarray}
&&\left.\frac{\partial ({\cal H}^{-1})^{ij}_{\nu\eta}}
{\partial \epsilon_{\alpha\beta}}\right|_{\fv}\!\!\!{\cal H}^{j\ell}_{\eta\kappa} + \!
({\cal H}^{-1})^{ij}_{\nu\eta}\left.\frac{\partial {\cal H}^{j\ell}_{\eta\kappa}}
{\partial \epsilon_{\alpha\beta}}\right|_{\fv}\!\! = \nonumber \\
&&\!\!\!\!\!\left.\frac{\partial ({\cal H}^{-1})^{ij}_{\nu\eta}}
{\partial \epsilon_{\alpha\beta}}\right|_{\fv}\!\!\!{\cal H}^{j\ell}_{\eta\kappa} + \!
({\cal H}^{-1})^{ij}_{\nu\eta}\left(\frac{\partial {\cal H}^{j\ell}_{\eta\kappa}}
{\partial \epsilon_{\alpha\beta}} +
{\cal V}^m_{\tau\alpha\beta}{\cal T}^{mj\ell}_{\tau\eta\kappa}
\right)\! = \!0 \nonumber .
\end{eqnarray}
From here
\begin{equation}
\left.\frac{\partial ({\cal H}^{-1})^{ij}_{\nu\eta}}
{\partial \epsilon_{\alpha\beta}}\right|_{\fv} \!\!\!\! = \!
-({\cal H}^{-1})^{i\ell}_{\nu\kappa}\!\!\left(\!
\frac{\partial {\cal H}^{\ell m}_{\kappa\chi}}{\partial \epsilon_{\alpha\beta}}
\!+\!{\cal V}^n_{\tau\alpha\beta}{\cal T}^{n\ell m}_{\tau\kappa\chi}\right)
\!\!({\cal H}^{-1})^{mj}_{\chi\eta}\ .
\end{equation}
With this relation and (\ref{nonAffineAcc}),
we carry out another constrained derivative of (\ref{zoo3}),
keeping only the most diverging terms
\begin{widetext}
\begin{equation}
-\left.\frac{\partial}{\partial \calBold{\epsilon} }\left(
\left[\calBold{T}\cdot\calBold{V}
\calBold{V}\right]\cdot\calBold{H}^{-1}\cdot
\left[\calBold{T}\cdot\calBold{V}\calBold{V}\right]\right)\right|_{\fv}
=
\calBold{T}\cdot\left[
\left(\calBold{H}^{-1}\cdot\left[\calBold{T}\cdot
\calBold{V}\calBold{V}\right]\right)\calBold{V}
\left(\calBold{H}^{-1}\cdot\left[\calBold{T}\cdot
\calBold{V}\calBold{V}\right]\right)\right] \nonumber
+ \mbox{14 symmetric terms}  \ .
\end{equation}
This leads us to the final expression for the instability strain $\gamma_P$:
\begin{equation}
\gamma_P = \gamma_0 - \frac{1}{2}\frac{
\left[\calBold{T}\cdot\calBold{V}_{xy}
\calBold{V}_{xy}\right]\cdot\calBold{H}^{-1}\cdot
\left[\calBold{T}\cdot\calBold{V}_{xy}\calBold{V}_{xy}\right]
}{
\calBold{T}\cdot\left[
\left(\calBold{H}^{-1}\cdot\left[\calBold{T}\cdot
\calBold{V}_{xy}\calBold{V}_{xy}\right]\right)\calBold{V}_{xy}
\left(\calBold{H}^{-1}\cdot\left[\calBold{T}\cdot
\calBold{V}_{xy}\calBold{V}_{xy}\right]\right)\right]
}\ .
\end{equation}
\end{widetext}
Notice that the above expression for $\gamma_P$ can be calculated
numerically by solving two linear equations; first, for $\calBold{V}$
using (\ref{nonAffVel}). With the solution for
$\calBold{V}$ in hand, one can then solve equation (\ref{foo1})
for $\left.\frac{\partial \calBold{V}}{\partial \calBold{\epsilon} }\right|_{\fv}
\simeq \calBold{H}^{-1}\cdot\left(\calBold{T}\cdot\calBold{V}\calBold{V}\right)$.
How these predictions work in practice can be read in \cite{10KLLP}.

%%%%%%%%%%%%%%%%%%%%%%%%%%%%%%%%%%%%%%%%%%%%%%%%%%%%%%%%%%%%%%%%
\section{summary and conclusions}
In this paper we derived closed-form expressions for the nonlinear elastic constants of
amorphous solids up to third order. We presented both the thermal and the athermal theory,
and demonstrated that the latter is obtained as a limit of the former when $T\to 0$. The
expressions derived above should be useful in numerical simulations where knowledge of
these nonlinear constants is indispensable due to the high values of the stresses obtained
near mechanical instabilities. In particular these expressions will allow extension of the evaluation
of these objects to local coarse-grained fields. Such an extension will be presented in a forthcoming
article. We demonstrated the use of these nonlinear objects in the context of understanding the
plasticity-induced anisotropy that arises in amorphous solids after a straining trajectory and in
predicting the plastic failure of amorphous solids to increasing strain.

\acknowledgments

This work had been supported in part by the Israel Science Foundation and the the Ministry of Science under
the French-Israeli collaboration. We are indebted to Anael Lemaitre for an extended discussion and collaboration
on these subjects. 

%%%%%%%%%%%%%%%%%%%%%%%%%%%%%%%%%%%%%%%%%%%%%%%%%%%%%%%%%
\appendix
%%%%%%%%%%%%%%%%%%%%%%%%%%%%%%%%%%%%%%%%%%%%%%%%%%%%%%%%%%%%%%%%%%%%%%%%%%%%%%%%%%%%%%%%%%%%%%%%%
\section{Constrained Derivatives}
\label{consder}
To provide an intuitive demonstration of the meaning of constrained derivatives,
consider two functions of two
independent variables $h(x,y)$ and $g(x,y)$. We want to understand, for instance,
how one should take
the partial derivative of $h$ with respect to $x$, keeping $g$ constant.
The constancy required from $g(x,y)$ means
that $y$ now depends on $x$, so that variations of $x$ \textbf{must} result in variations in $y$, namely
\begin{equation}\label{moo6}
dy = \left. \frac{\partial y}{\partial x}\right|_gdx\ .
\end{equation}
We write the total variation of $h$ as
\begin{equation}
dh = \left. \frac{\partial h}{\partial x}\right|_y dx
+ \left. \frac{\partial h}{\partial y}\right|_x dy \ .
\end{equation}
We now impose the dependence of the variation in $y$ on the variation in $x$:
\begin{equation}
dh = \left. \frac{\partial h}{\partial x}\right|_y dx
+ \left. \frac{\partial h}{\partial y}\right|_x \left. \frac{\partial y}{\partial x}\right|_gdx\ .
\end{equation}
From here
\begin{equation}\label{moo7}
\left. \frac{\partial h}{\partial x}\right|_g =
\left. \frac{\partial h}{\partial x}\right|_y
+ \left. \frac{\partial h}{\partial y}\right|_x \left. \frac{\partial y}{\partial x}\right|_g\ .
\end{equation}
It is important to notice that in this framework, constrained
partial derivatives do not necessarily commute;
to exemplify this issue, consider the constraint $g(x,y) = x^2 + y^2 = 1$
in the range of positive $x$ and $y$.
The partial derivative of $y$ with respect to $x$ at constant $g$ is
\begin{equation}
\left.\frac{\partial y}{\partial x}\right|_g = -\frac{x}{\sqrt{1-x^2}}\ .
\end{equation}
We first consider
\begin{equation}
\left.\frac{\partial^2y}{\partial x^2}\right|_{g,y} \!\!=\!
-\left.\frac{\partial}{\partial x}\right|_y\!\left(\frac{x}{\sqrt{1-x^2}}\right) =
-\left.\frac{\partial}{\partial x}\right|_y\!\!\left(\frac{x}{y}\right) = -\frac{1}{y}\ .
\end{equation}
Next, it is immediate that
\begin{equation}
\left.\frac{\partial^2y}{\partial x^2}\right|_{y,g} =
\left.\frac{\partial }{\partial x}\right|_{g} \left.\frac{\partial y}{\partial x}\right|_{y} =  0\ ,
\end{equation}
since obviously any variation of $y$ keeping $y$ constant is zero. So, generally
\begin{equation}
\left.\frac{\partial^2y}{\partial x^2}\right|_{y,g} \ne
\left.\frac{\partial^2y}{\partial x^2}\right|_{g,y} \ .
\end{equation}

%%%%%%%%%%%%%%%%%%%%%%%%%%%%%%%%%%%%%%%%%%%%%%%%%%%%%%%%%%%%%%%%%%%%%%%%%%%%%%%
\section{Kinetic energy derivatives}
\label{Kinender}
We finally derive expressions for kinetic energy derivatives that appear in (\ref{E_terms}).
Given a transformation of coordinates ${\Hv}$, the transformation of momenta is
dictated by requiring that the transformation is canonical, namely that
\begin{equation}
\frac{\partial x_\alpha' }{\partial x_\nu}\frac{\partial p_\beta' }{\partial p_\nu} -
\frac{\partial x_\alpha' }{\partial p_\nu}\frac{\partial p_\beta' }{\partial x_\nu}
= \delta_{\alpha\beta}\ ,\label{poisson}
\end{equation}
where prime denotes transformed coordinates and momenta,
and in this chapter repeated indices are summed over.
Assume that $p_\beta' = A_{\beta\kappa}p_\kappa$, such that
$\frac{\partial p_\beta'}{\partial p_\nu} = A_{\beta\kappa}\delta_{\kappa\nu}$.
Inserting this in equation (\ref{poisson}):
\begin{equation}
H_{\alpha\mu}\delta_{\nu\mu}A_{\beta\kappa}\delta_{\kappa\nu} =
H_{\alpha\nu}A_{\beta\nu} = H_{\alpha\nu}A^T_{\nu\beta} = \delta_{\alpha\beta}\ .
\end{equation}
This means that $\Hv \Av^T = \Iv$, or $\Av = (\Hv^{-1})^T$,
such that the transformation of the momenta is
\begin{equation}
p_\alpha' = (H^{-1})^T_{\alpha\beta}p_\beta\ .
\end{equation}
The kinetic energy after imposing a small strain is
\begin{equation}\label{moo2}
\begin{split}
K = \sFrac{1}{2}p'^{i}_\alpha p'^{i}_\alpha = &\
\sFrac{1}{2} (H^{-1})^T_{\alpha\nu}p^i_\nu (H^{-1})^T_{\alpha\beta}p^i_\beta \\
= &\  \sFrac{1}{2} H^{-1}_{\nu\alpha}(H^{-1})^T_{\alpha\beta}p^i_\nu p^i_\beta \ .
\end{split}\end{equation}
From (\ref{strainT}) we have
$\Hv^T\Hv = 2\mathBold{\epsilon} + \Iv$, hence
\begin{equation}
\Hv^{-1}(\Hv^{-1})^T  = (\Hv^T\Hv)^{-1} = (2\mathBold{\epsilon} + \Iv)^{-1}\ .
\end{equation}
Expanding this up to $\mathBold{\epsilon}^3$:
\begin{equation}
(2\mathBold{\epsilon} + \Iv)^{-1} \simeq \Iv - 2\mathBold{\epsilon}
+ 4\mathBold{\epsilon}^2 -8\mathBold{\epsilon}^3 + {\cal O}(\mathBold{\epsilon}^4)\ .
\end{equation}
Inserting this back to (\ref{moo2}):
\begin{equation}\begin{split}
K \simeq & \ \sFrac{1}{2}\delta_{\alpha\beta}p^i_\alpha p^i_\beta
-\epsilon_{\alpha\beta}p^i_\alpha p^i_\beta +
2\epsilon_{\alpha\nu}\epsilon_{\nu\beta}p^i_\alpha p^i_\beta \\ &\
-4\epsilon_{\alpha\nu}\epsilon_{\nu\eta}\epsilon_{\eta\beta}p^i_\alpha p^i_\beta
+ {\cal O}(\epsilon^4)\ .
\end{split}\end{equation}
We can now calculate
\begin{equation}\label{firstDerivativeKinetic}
\begin{split}
\frac{\partial K}{\partial \epsilon_{\alpha\beta}} = & -p^i_\alpha p^i_\beta
 +2p^i_\alpha p^i_\nu\epsilon_{\nu\beta}
+2p^i_\beta p^i_\nu\epsilon_{\nu\alpha} \\
& -4\epsilon_{\beta\nu}\epsilon_{\nu\eta}p^i_\eta p^i_\alpha
-4\epsilon_{\alpha\nu}\epsilon_{\beta\eta}p^i_\nu p^i_\eta
-4\epsilon_{\nu\eta}\epsilon_{\eta\alpha}p^i_\nu p^i_\beta \ .
\end{split}
\end{equation}
The second (symmetrized) derivative is
\begin{widetext}
\begin{equation}\label{secondDerivativeKinetic}
\begin{split}
\frac{\partial^2 K}{\partial\epsilon_{\nu\eta}\partial \epsilon_{\alpha\beta}} =\  &
 \delta_{\beta\eta} p^i_\alpha p^i_\nu
+\delta_{\alpha\eta} p^i_\beta p^i_\nu
+ \delta_{\beta\nu} p^i_\alpha p^i_\eta
+\delta_{\alpha\nu} p^i_\beta p^i_\eta\\&
-2\delta_{\beta\nu} \epsilon_{\eta\theta}p^i_\theta p^i_\alpha
-2 \epsilon_{\beta\nu}p^i_\eta p^i_\alpha
-2\delta_{\beta\eta} \epsilon_{\nu\theta}p^i_\theta p^i_\alpha
-2 \epsilon_{\beta\eta}p^i_\nu p^i_\alpha \\
&
-2\delta_{\alpha\nu} \epsilon_{\beta\theta}p^i_\eta p^i_\theta
-2\delta_{\beta\nu} \epsilon_{\alpha\theta}p^i_\theta p^i_\eta
-2\delta_{\alpha\eta} \epsilon_{\beta\theta}p^i_\nu p^i_\theta
-2\delta_{\beta\eta} \epsilon_{\alpha\theta}p^i_\theta p^i_\nu\\
&
-2 \epsilon_{\eta\alpha}p^i_\nu p^i_\beta
-2\delta_{\alpha\eta} \epsilon_{\theta\nu}p^i_\theta p^i_\beta
-2 \epsilon_{\nu\alpha}p^i_\eta p^i_\beta
-2\delta_{\alpha\nu} \epsilon_{\theta\eta}p^i_\theta p^i_\beta\ ,
\end{split}
\end{equation}
and the third (symmetrized) derivative is
\begin{equation}\label{thirdDerivativeKinetic}
\begin{split}
\frac{\partial^3 K}{\partial\epsilon_{\kappa\chi}
\partial\epsilon_{\nu\eta}\partial \epsilon_{\alpha\beta}} = &
-\delta_{\beta\nu}\delta_{\kappa\eta} p^i_\chi p^i_\alpha
-\delta_{\kappa\beta}\delta_{\chi\nu} p^i_\eta p^i_\alpha
-\delta_{\beta\eta}\delta_{\kappa\nu} p^i_\chi p^i_\alpha
-\delta_{\kappa\beta}\delta_{\chi\eta} p^i_\nu p^i_\alpha \\
&
-\delta_{\beta\nu}\delta_{\chi\eta} p^i_\kappa p^i_\alpha
-\delta_{\chi\beta}\delta_{\kappa\nu} p^i_\eta p^i_\alpha
-\delta_{\beta\eta}\delta_{\chi\nu} p^i_\kappa p^i_\alpha
-\delta_{\chi\beta}\delta_{\kappa\eta} p^i_\nu p^i_\alpha \\
&
-\delta_{\alpha\nu}\delta_{\kappa\beta} p^i_\eta p^i_\chi
-\delta_{\beta\nu}\delta_{\kappa\alpha} p^i_\chi p^i_\eta
-\delta_{\alpha\eta}\delta_{\kappa\beta} p^i_\nu p^i_\chi
-\delta_{\beta\eta}\delta_{\kappa\alpha} p^i_\chi p^i_\nu\\
&
-\delta_{\alpha\nu}\delta_{\chi\beta} p^i_\eta p^i_\kappa
-\delta_{\beta\nu}\delta_{\chi\alpha} p^i_\kappa p^i_\eta
-\delta_{\alpha\eta}\delta_{\chi\beta} p^i_\nu p^i_\kappa
-\delta_{\beta\eta}\delta_{\chi\alpha} p^i_\kappa p^i_\nu\\
&
-\delta_{\kappa\eta}\delta_{\chi\alpha} p^i_\nu p^i_\beta
-\delta_{\alpha\eta}\delta_{\chi\nu} p^i_\kappa p^i_\beta
-\delta_{\kappa\nu}\delta_{\chi\alpha} p^i_\eta p^i_\beta
-\delta_{\alpha\nu}\delta_{\chi\eta} p^i_\kappa p^i_\beta \\
&
-\delta_{\chi\eta}\delta_{\kappa\alpha} p^i_\nu p^i_\beta
-\delta_{\alpha\eta}\delta_{\kappa\nu} p^i_\chi p^i_\beta
-\delta_{\chi\nu}\delta_{\kappa\alpha} p^i_\eta p^i_\beta
-\delta_{\alpha\nu}\delta_{\kappa\eta} p^i_\chi p^i_\beta\ .
\end{split}
\end{equation}
\end{widetext}

\section{potential energy derivatives for pairwise potentials}
\label{potentialEnergyDerivatives}
In this appendix we calculate all partial derivatives of the potential energy with respect to
strain and particle positions, of all required orders.
In the following section, partial derivatives with respect
to coordinates $\xv$ should be understood as taken at constant $\calBold{\epsilon}$, and
partial derivatives with respect to strain $\calBold{\epsilon}$ should be understood as taken
at constant non-affine fields $\Xv$ (no relaxation allowed for).

We first carry out partial derivatives of the potential energy with respect to
particle coordinates. The potential energy is given by
\begin{equation}
U = \sFrac{1}{2}\sum_{i\ne j} \phi^{ij}\ ,
\end{equation}
where $\phi^{ij}$ is the pairwise interaction potential between
the $i$'th and $j$'th particles.
The negative of the forces are given by
\begin{equation}
-f^\ell_\alpha = \frac{\partial U}{\partial x^\ell_\alpha}
= \sFrac{1}{2}\sum_{i\ne j} \frac{\partial \phi^{ij}}{\partial x^\ell_\alpha}
= \sFrac{1}{2}\sum_{i\ne j} \frac{\partial \phi^{ij}}{\partial r^{ij}}
\frac{\partial r^{ij}}{\partial x^\ell_\alpha}\ .
\end{equation}
Since $r^{ij} = \sqrt{(x^j_\beta - x^i_\beta)(x^j_\beta - x^i_\beta)}$ then
$\frac{\partial r^{ij}}{\partial x^\ell_\alpha} = \frac{r^{ij}_\alpha}{r^{ij}}(\delta^{j\ell}
- \delta^{i\ell})$. With the notations
$\phi_r, \phi_{rr}$ and $\phi_{rrr}$ for the first, second and third derivatives of $\phi(r)$
with respect to $r$, respectively, we now have
\begin{widetext}
\begin{equation}
\frac{\partial U}{\partial x^\ell_\alpha}  =
\sFrac{1}{2}\sum_{i\ne j}\phi^{ij}_r
\frac{r^{ij}_\alpha}{r^{ij}}(\delta^{j\ell} - \delta^{i\ell})  =
\sFrac{1}{2}\sum_{i\ne \ell}\phi^{i\ell}_r
\frac{r^{i\ell}_\alpha}{r^{i\ell}}-\sFrac{1}{2}\sum_{\ell\ne j}\phi^{\ell j}_r
\frac{r^{\ell j}_\alpha}{r^{\ell j}}  =
\sFrac{1}{2}\sum_{i\ne \ell}\phi^{i\ell}_r
\frac{r^{i\ell}_\alpha}{r^{i\ell}}
+\sFrac{1}{2}\sum_{j\ne\ell}\phi^{j\ell}_r
\frac{r^{j\ell}_\alpha}{r^{j\ell}}
= \sum_{i\ne \ell}
\frac{\phi^{i\ell}_rr^{i\ell}_\alpha}{r^{i\ell}}.
\end{equation}
The Hessian is defined as
\begin{eqnarray}
{\cal{H}}^{ij}_{\alpha\beta} =
\frac{\partial^2 U }{\partial x^{j}_{\beta} \partial x^{i}_{\alpha}}
& = & \frac{1}{2}\frac{\partial }{\partial x^{j}_{\beta}}
\sum_{k,\ell} \frac{\partial \phi^{k\ell}}
{\partial r^{k\ell}}\frac{\partial r^{k\ell}}{\partial x^{i}_{\alpha}}
 =  \frac{1}{2}\frac{\partial }{\partial x^{j}_{\beta}}
\sum_{k,\ell} \frac{\phi^{k\ell}_r}
{r^{k\ell}}r^{k\ell}_\alpha
\left( \delta^{i\ell} - \delta^{ik}\right)
 =  \frac{\partial }{\partial x^{j}_{\beta}}
\sum_{k \ne i} \frac{\phi^{ki}_r}
{r^{ki}}r^{ki}_\alpha\  \nonumber \\
& = & \sum_{k \ne i}
\left(\frac{\phi_{rr}^{ik}}{(r^{ik})^2} -\frac{\phi_r^{ik}}{(r^{ik})^3}\right)
r^{ik}_\alpha r^{ik}_\beta(\delta^{ij} - \delta^{kj}) +
\delta_{\alpha\beta}\sum_{k \ne i} \frac{\phi^{ki}_r}
{r^{ki}}(\delta^{ij} - \delta^{kj})\ .
\end{eqnarray}
The off diagonal term $i\ne j$ of the Hessian is given by
\begin{equation}\label{offDiagonalHessian}
{\cal{H}}^{ij}_{\alpha\beta} = -\left(\frac{\phi^{ij}_{rr}}{(r^{ij})^2}
- \frac{\phi^{ij}_r}{(r^{ij})^3}\right)
r^{ij}_\alpha r^{ij}_\beta
- \delta_{\alpha\beta}\frac{\phi^{ij}_r}{r^{ij}}\ ,
\end{equation}
and the ${\cal{H}}^{ii}_{\alpha\beta}$ is given by
\begin{equation}\label{diagonalHessian}
{\cal {H}}^{ii}_{\alpha\beta} = -\sum_{j \ne i} {\cal {H}}^{ij}_{\alpha\beta}\ .
\end{equation}

We now calculate $\frac{\partial ^3U}{\partial x^\ell_\chi \partial x^j_\beta \partial x^i_\alpha}$;
assuming first $i\ne j$, we take a derivative of Eq.~(\ref{offDiagonalHessian})
\begin{equation}
\frac{\partial {\cal {H}}^{ij}_{\alpha\beta}}{\partial x^\ell_\chi} =
-(\delta^{j\ell} - \delta^{i\ell})\!\!\left[\!\left( \frac{\phi_{rrr}^{ij}}{(r^{ij})^3} \!-\!
\frac{3\phi_{rr}^{ij}}{(r^{ij})^4}
\!+\! \frac{3\phi_{r}^{ij}}{(r^{ij})^5}\right)r_\alpha^{ij}r_\beta^{ij}r_\chi^{ij} + \left(
\frac{\phi_{rr}^{ij}}{(r^{ij})^2} \!-\! \frac{\phi_r^{ij}}{(r^{ij})^3}
\right)\!\!\left(\delta_{\alpha\chi}r_\beta^{ij} \!+\! \delta_{\beta\chi}r_\alpha^{ij}
\!+\! \delta_{\alpha\beta}r_\chi^{ij}\right)\!\right].
\end{equation}
If $i\ne j\ne\ell$ then $\frac{\partial ^3U}{\partial x^\ell_\chi \partial x^j_\beta \partial x^i_\alpha} = 0$,
so in the above expression $\ell$ must be equal to either $i$ or $j$ (but not to the both).
Due to the symmetry
$\frac{\partial ^3U}{\partial x^i_\chi \partial x^j_\beta \partial x^i_\alpha} =
\frac{\partial ^3U}{\partial x^i_\chi \partial x^i_\alpha \partial x^j_\beta} =
\frac{\partial ^3U}{\partial x^j_\beta \partial x^i_\chi \partial x^i_\alpha}$, we can
limit the discussion to two cases; first, if two of the particle indices are equal,
and different from the third, then
\begin{equation}
\frac{\partial ^3U}{\partial x^i_\chi \partial x^j_\beta \partial x^i_\alpha} = \!
\left[\!\left( \frac{\phi_{rrr}^{ij}}{(r^{ij})^3} \!-\!
\frac{3\phi_{rr}^{ij}}{(r^{ij})^4}
\!+\! \frac{3\phi_{r}^{ij}}{(r^{ij})^5}\right)r_\alpha^{ij}r_\beta^{ij}r_\chi^{ij} + \left(
\frac{\phi_{rr}^{ij}}{(r^{ij})^2} \!-\! \frac{\phi_r^{ij}}{(r^{ij})^3}
\right)\!\!\left(\delta_{\alpha\chi}r_\beta^{ij} \!+\! \delta_{\beta\chi}r_\alpha^{ij}
\!+\! \delta_{\alpha\beta}r_\chi^{ij}\right)\!\right],
\end{equation}
where we assumed $i\ne j$. The diagonal term is then
\begin{equation}
\frac{\partial ^3U}{\partial x^i_\chi \partial x^i_\beta \partial x^i_\alpha} =
-\sum_{j\ne i}\frac{\partial ^3U}{\partial x^i_\chi \partial x^j_\beta \partial x^i_\alpha}\ .
\end{equation}

We now turn to derivatives with respect to strain $\mathBold{\epsilon}$;
consider first the {\bf change} in distance $\delta r^{ij}$ between particles $i$ and $j$,
before and after imposing a deformation.
For a given imposed deformation represented by $\Hv$,
each component of the pairwise distance transforms via
$r^{ij}_\alpha \to H_{\alpha\beta}r^{ij}_\alpha$,
so the distance between the pair $(i,j)$ {\bf after} the deformation is
$\sqrt{H^T_{\alpha\nu}H_{\nu\beta}r^{ij}_\alpha r^{ij}_\beta}$.
Since by definition $\Hv^T\Hv = 2\mathBold{\epsilon} + \Iv$, the change in the pairwise distance is
\begin{equation}\begin{split}\label{foo2}
\delta r^{ij} =  &\, \sqrt{H^T_{\alpha\nu}
H_{\nu\beta}r^{ij}_\alpha r^{ij}_\beta} - r^{ij}
= \sqrt{(2\epsilon_{\alpha\beta}+ \delta_{\alpha\beta})r^{ij}_\alpha r^{ij}_\beta} - r^{ij}
=  r^{ij}\sqrt{1 +
\frac{2\epsilon_{\alpha\beta}r^{ij}_\alpha r^{ij}_\beta}{r^{ij}_\nu r^{ij}_\nu}}- r^{ij}  \\
\simeq &\, r^{ij}\left(1 + \frac{\epsilon_{\alpha\beta}r^{ij}_\alpha r^{ij}_\beta}{(r^{ij})^2}
- \frac{1}{2}\frac{\epsilon_{\alpha\beta}r^{ij}_\alpha r^{ij}_\beta
\epsilon_{\nu\eta}r^{ij}_\nu r^{ij}_\eta}{(r^{ij})^4}
+\frac{1}{2}\frac{\epsilon_{\alpha\beta}r^{ij}_\alpha r^{ij}_\beta
\epsilon_{\nu\eta}r^{ij}_\nu
r^{ij}_\eta\epsilon_{\kappa\chi}r^{ij}_\kappa r^{ij}_\chi}{(r^{ij})^6}\right) - r^{ij} \\
= &\, \frac{\epsilon_{\alpha\beta}r^{ij}_\alpha r^{ij}_\beta}{r^{ij}}
- \frac{1}{2}\frac{\epsilon_{\alpha\beta}r^{ij}_\alpha r^{ij}_\beta
\epsilon_{\nu\eta}r^{ij}_\nu r^{ij}_\eta}{(r^{ij})^3}
+\frac{1}{2}\frac{\epsilon_{\alpha\beta}r^{ij}_\alpha r^{ij}_\beta
\epsilon_{\nu\eta}r^{ij}_\nu r^{ij}_\eta
\epsilon_{\kappa\chi}r^{ij}_\kappa r^{ij}_\chi}{(r^{ij})^5}\ .
\end{split}\end{equation}
We write the potential energy as a sum of pairwise contributions $U = \sum_{i<j}\phi^{ij}$,
and expand the pairwise potential $\phi^{ij}$ in terms of $\delta r^{ij}$:
\begin{equation}
\phi^{ij} = \phi_0^{ij} + \phi^{ij}_r\delta r^{ij} +
\sFrac{1}{2}\phi^{ij}_{rr}\delta r^{ij}\delta r^{ij} +
\sFrac{1}{6}\phi^{ij}_{rrr}\delta r^{ij}\delta r^{ij}\delta r^{ij} + \ldots\ ,
\end{equation}
Plugging (\ref{foo2}) in the above expansion gives $\phi^{ij}$ in terms of $\mathBold{\epsilon}$,
keeping terms up to $\mathBold{\epsilon}^3$:
\begin{equation}\label{phiOfEpsilon}
\begin{split}
\phi^{ij} = & \phi^{ij}_0 +
\epsilon_{\alpha\beta}\frac{\phi_r^{ij}r^{ij}_\alpha r^{ij}_\beta}{r^{ij}} +
\sFrac{1}{2}\epsilon_{\alpha\beta}\epsilon_{\nu\eta}
r^{ij}_\alpha r^{ij}_\beta r^{ij}_\nu r^{ij}_\eta\left(
\frac{\phi_{rr}^{ij}}{(r^{ij})^2} -\frac{\phi_r^{ij}}{(r^{ij})^3}\right) \\
& + \,\,\sFrac{1}{6}\epsilon_{\alpha\beta}\epsilon_{\nu\eta}\epsilon_{\kappa\chi}
r^{ij}_\alpha r^{ij}_\beta r^{ij}_\nu r^{ij}_\eta r^{ij}_\kappa r^{ij}_\chi
\left(\frac{3\phi^{ij}_{r}}{(r^{ij})^5} - \frac{3\phi^{ij}_{rr}}{(r^{ij})^4}
+\frac{\phi^{ij}_{rrr}}{(r^{ij})^3} \right)\ .
\end{split}
\end{equation}
From here the derivatives of $\phi$ with respect to strain can be calculated as
\begin{equation}\label{d_phi_d_epsilon}
\frac{\partial \phi^{ij}}{\partial \epsilon_{\alpha\beta}} =
\frac{\phi_r^{ij}r^{ij}_\alpha r^{ij}_\beta}{r^{ij}} +
\epsilon_{\nu\eta}r^{ij}_\alpha r^{ij}_\beta r^{ij}_\nu r^{ij}_\eta\left(
\frac{\phi_{rr}^{ij}}{(r^{ij})^2} -\frac{\phi_r^{ij}}{(r^{ij})^3}\right)
+\sFrac{1}{2}\epsilon_{\nu\eta}\epsilon_{\kappa\chi}
r^{ij}_\alpha r^{ij}_\beta r^{ij}_\nu r^{ij}_\eta r^{ij}_\kappa r^{ij}_\chi
\left(\frac{3\phi^{ij}_{r}}{(r^{ij})^5}
-\frac{3\phi^{ij}_{rr}}{(r^{ij})^4}+\frac{\phi^{ij}_{rrr}}{(r^{ij})^3} \right) ,
\end{equation}
\begin{equation}\label{d2_phi_d_epsilon2}
\frac{\partial^2 \phi^{ij}}{\partial \epsilon_{\nu\eta}
\partial \epsilon_{\alpha\beta}} =
r^{ij}_\alpha r^{ij}_\beta r^{ij}_\nu r^{ij}_\eta\left(
\frac{\phi_{rr}^{ij}}{(r^{ij})^2} -\frac{\phi_r^{ij}}{(r^{ij})^3}\right)
+\epsilon_{\kappa\chi}
r^{ij}_\alpha r^{ij}_\beta r^{ij}_\nu r^{ij}_\eta r^{ij}_\kappa r^{ij}_\chi
\left(\frac{3\phi^{ij}_{r}}{(r^{ij})^5} - \frac{3\phi^{ij}_{rr}}{(r^{ij})^4}
+\frac{\phi^{ij}_{rrr}}{(r^{ij})^3} \right)\ ,
\end{equation}
and
\begin{equation}\label{d3_phi_d_epsilon3}
\frac{\partial^3 \phi^{ij}}{\partial\epsilon_{\kappa\chi}
\partial \epsilon_{\nu\eta}\partial \epsilon_{\alpha\beta}} =
r^{ij}_\alpha r^{ij}_\beta r^{ij}_\nu r^{ij}_\eta r^{ij}_\kappa r^{ij}_\chi
\left(\frac{3\phi^{ij}_{r}}{(r^{ij})^5} - \frac{3\phi^{ij}_{rr}}{(r^{ij})^4}
+\frac{\phi^{ij}_{rrr}}{(r^{ij})^3} \right)
\end{equation}
The derivatives with respect to strain of the total potential energy at
$\mathBold{\epsilon} = 0$ are given by:
\begin{equation}
\left. \frac{\partial U}{\partial \epsilon_{\alpha\beta}}
\right|_{\mathBold{\epsilon}=0} =
\sum_{i<j}\frac{\phi_r^{ij}r^{ij}_\alpha r^{ij}_\beta}{r^{ij}} \ , \quad
\left.\frac{\partial^2 U}{\partial \epsilon_{\nu\eta}\partial \epsilon_{\alpha\beta}}
\right|_{\mathBold{\epsilon} = 0} =
\sum_{i<j}r^{ij}_\alpha r^{ij}_\beta r^{ij}_\nu r^{ij}_\eta\left(
\frac{\phi_{rr}^{ij}}{(r^{ij})^2} -\frac{\phi_r^{ij}}{(r^{ij})^3}\right)\ ,
\end{equation}
and
\begin{equation}\label{thirdDerivativeU}
\left. \frac{\partial^3 U}{\partial\epsilon_{\kappa\chi}
\partial \epsilon_{\nu\eta}\partial \epsilon_{\alpha\beta}}\right|_{\mathBold{\epsilon}=0}
=\sum_{i<j}r^{ij}_\alpha r^{ij}_\beta r^{ij}_\nu r^{ij}_\eta r^{ij}_\kappa r^{ij}_\chi
\left(\frac{3\phi^{ij}_{r}}{(r^{ij})^5} - \frac{3\phi^{ij}_{rr}}{(r^{ij})^4}
+\frac{\phi^{ij}_{rrr}}{(r^{ij})^3} \right)
\end{equation}

Turning now to the mixed derivatives, we first consider
$\frac{\partial^2\phi^{ij}}{\partial x^k_\nu \partial \epsilon_{\alpha\beta}}$;
following (\ref{d_phi_d_epsilon}), we obtain
\begin{equation}\label{d2_phi_d_epsilon_dx}
\left. \frac{\partial^2\phi^{ij}}{\partial x^k_\nu \partial \epsilon_{\alpha\beta}}
\right|_{\mathBold{\epsilon} = 0}
= (\delta^{jk} - \delta^{ik})\left[
\left(\frac{\phi_{rr}^{ij}}{(r^{ij})^2} -\frac{\phi_r^{ij}}{(r^{ij})^3}\right)
r^{ij}_\nu r^{ij}_\alpha r^{ij}_\beta +
\frac{\phi^{ij}_r}{r^{ij}}(\delta_{\nu\alpha}r^{ij}_\beta +
\delta_{\nu\beta}r^{ij}_\alpha)\right]\ .
\end{equation}
Next we calculate the mixed derivative
$\frac{\partial^3 \phi^{ij}}
{\partial x^k_\nu\partial \epsilon_{\nu\eta}\partial \epsilon_{\alpha\beta}}$;
following (\ref{d2_phi_d_epsilon2}), we obtian
\begin{equation}\label{d3_phi_d_epsilon2_dx}
\begin{split}
\left.\frac{\partial^3 \phi^{ij}}
{\partial x^k_\chi\partial \epsilon_{\nu\eta}\partial \epsilon_{\alpha\beta}}
\right|_{\mathBold{\epsilon} = 0}
= & \ (\delta^{jk} - \delta^{ik})\left[ \left(\frac{\phi_{rrr}^{ij}}{(r^{ij})^3} -
\frac{3\phi_{rr}^{ij}}{(r^{ij})^4} + \frac{3\phi_{r}^{ij}}{(r^{ij})^5}\right)
r^{ij}_\chi r^{ij}_\alpha r^{ij}_\beta r^{ij}_\nu r^{ij}_\eta \right. \\
&\left.  + \left(
\delta_{\alpha\chi}r^{ij}_\beta r^{ij}_\nu r^{ij}_\eta +
\delta_{\beta\chi}r^{ij}_\alpha r^{ij}_\nu r^{ij}_\eta +
\delta_{\nu\chi}r^{ij}_\alpha r^{ij}_\beta r^{ij}_\eta +
\delta_{\eta\chi}r^{ij}_\alpha r^{ij}_\beta r^{ij}_\nu \right)
\left(\frac{\phi_{rr}^{ij}}{(r^{ij})^2} -\frac{\phi_r^{ij}}{(r^{ij})^3}\right)
\right]
\end{split}
\end{equation}

The last mixed derivative is
$\frac{\partial^3 \phi^{ij}}
{\partial x^\ell_\chi \partial x^k_\nu \partial \epsilon_{\alpha\beta}}$;
following Eq.~(\ref{d2_phi_d_epsilon_dx}), we obtain
\begin{equation}\label{d3_phi_d_epsilon_dx2}\begin{split}
\left.\frac{\partial^3 \phi^{ij}}
{\partial x^\ell_\chi \partial x^k_\nu \partial \epsilon_{\alpha\beta}}
\right|_{\mathBold{\epsilon} = 0}=  &
(\delta^{jk} - \delta^{ik})(\delta^{j\ell} - \delta^{i\ell}) \left\{
\left(\frac{\phi_{rrr}^{ij}}{(r^{ij})^3} -
\frac{3\phi_{rr}^{ij}}{(r^{ij})^4} + \frac{3\phi_{r}^{ij}}{(r^{ij})^5}\right)
r^{ij}_\alpha r^{ij}_\beta r^{ij}_\nu r^{ij}_\chi  \right. \\
&+\left(\frac{\phi_{rr}^{ij}}{(r^{ij})^2} -\frac{\phi_r^{ij}}{(r^{ij})^3}\right)
\left(
\delta_{\alpha\chi}r^{ij}_\beta r^{ij}_\nu +
\delta_{\beta\chi}r^{ij}_\alpha r^{ij}_\nu +
\delta_{\nu\chi}r^{ij}_\alpha r^{ij}_\beta +
\delta_{\nu\alpha}r^{ij}_\beta r^{ij}_\chi +
\delta_{\nu\beta}r^{ij}_\alpha r^{ij}_\chi
\right) \\
&+\frac{\phi_r^{ij}}{r^{ij}}\left(
\delta_{\nu\alpha}\delta_{\beta\chi} +
\delta_{\nu\beta}\delta_{\alpha\chi}
\right)
\left. \vphantom{\frac{\sum}{}}\right\}\ .
\end{split}\end{equation}

For any quantity $A^{ij}$ such that $A^{ij} = -A^{ji}$, we have
\[
\sum_{i<j}(\delta^{jk} - \delta^{ik})A^{ij} =
\sFrac{1}{2}\sum_{i}\sum_{j\ne i}(\delta^{jk} - \delta^{ik})A^{ij} =
\sFrac{1}{2}\sum_{i\ne k}A^{ik} -
\sFrac{1}{2}\sum_{j\ne k}A^{kj} =
\sum_{i\ne k}A^{ik}\ .
\]
Since in equations (\ref{d2_phi_d_epsilon_dx}), (\ref{d3_phi_d_epsilon2_dx}) and
(\ref{d3_phi_d_epsilon_dx2})
the terms that correspond to $A^{ij}$ in the above equation are
anti-symmetric in $\{i,j\}$, so we can directly write the results for the mixed
energy derivatives:
\begin{equation}\label{d2_U_d_epsilon_dx}
\left.\frac{\partial^2U}{\partial x^k_\nu \partial \epsilon_{\alpha\beta}}
\right|_{\mathBold{\epsilon} = 0}
= \sum_{i\ne k}\left[
\left(\frac{\phi_{rr}^{ik}}{(r^{ik})^2} -\frac{\phi_r^{ik}}{(r^{ik})^3}\right)
r^{ik}_\nu r^{ik}_\alpha r^{ik}_\beta +
\frac{\phi^{ik}_r}{r^{ik}}(\delta_{\nu\alpha}r^{ik}_\beta +
\delta_{\nu\beta}r^{ik}_\alpha)\right]\ ,
\end{equation}
\begin{equation}
\begin{split}
\left.\frac{\partial^3 U}
{\partial x^k_\chi\partial \epsilon_{\nu\eta}\partial \epsilon_{\alpha\beta}}
\right|_{\mathBold{\epsilon} = 0}= & \
\sum_{i\ne k}\left[ \left(\frac{\phi_{rrr}^{ik}}{(r^{ik})^3} -
\frac{3\phi_{rr}^{ik}}{(r^{ik})^4} + \frac{3\phi_{r}^{ik}}{(r^{ik})^5}\right)
r^{ik}_\chi r^{ik}_\alpha r^{ik}_\beta r^{ik}_\nu r^{ik}_\eta \right. \\
&\hspace{-0.2cm} \left.  +
\left(
\delta_{\alpha\chi}r^{ij}_\beta r^{ij}_\nu r^{ij}_\eta +
\delta_{\beta\chi}r^{ij}_\alpha r^{ij}_\nu r^{ij}_\eta +
\delta_{\nu\chi}r^{ij}_\alpha r^{ij}_\beta r^{ij}_\eta +
\delta_{\eta\chi}r^{ij}_\alpha r^{ij}_\beta r^{ij}_\nu \right)
\left(\frac{\phi_{rr}^{ik}}{(r^{ik})^2} -\frac{\phi_r^{ik}}{(r^{ik})^3}\right)
\right]\ ,
\end{split}
\end{equation}
and
\begin{equation}\begin{split}
\left.\frac{\partial^3 U}
{\partial x^\ell_\chi \partial x^k_\nu \partial \epsilon_{\alpha\beta}}
\right|_{\mathBold{\epsilon} = 0}=  &
\sum_{i\ne k}(\delta^{k\ell} - \delta^{i\ell}) \left\{
\left(\frac{\phi_{rrr}^{ik}}{(r^{ik})^3} -
\frac{3\phi_{rr}^{ik}}{(r^{ik})^4} + \frac{3\phi_{r}^{ik}}{(r^{ik})^5}\right)
r^{ik}_\alpha r^{ik}_\beta r^{ik}_\nu r^{ik}_\chi  \right. \\
&+\left(\frac{\phi_{rr}^{ik}}{(r^{ik})^2} -\frac{\phi_r^{ik}}{(r^{ik})^3}\right)
\left(
\delta_{\alpha\chi}r^{ik}_\beta r^{ik}_\nu \!+\!
\delta_{\beta\chi}r^{ik}_\alpha r^{ik}_\nu \!+\!
\delta_{\nu\chi}r^{ik}_\alpha r^{ik}_\beta \!+\!
\delta_{\nu\alpha}r^{ik}_\beta r^{ik}_\chi \!+\!
\delta_{\nu\beta}r^{ik}_\alpha r^{ik}_\chi
\right) \\
&+\frac{\phi_r^{ik}}{r^{ik}}\left(
\delta_{\nu\alpha}\delta_{\beta\chi} +
\delta_{\nu\beta}\delta_{\alpha\chi}
\right)
\left. \vphantom{\frac{\sum}{}}\right\}\ .
\end{split}\end{equation}
In the case for which $\ell \ne k$, this is
\begin{equation}\label{d3_U_d_epsilon_dx2}\begin{split}
\left.\frac{\partial^3 U}
{\partial x^\ell_\chi \partial x^k_\nu \partial \epsilon_{\alpha\beta}}
\right|_{\mathBold{\epsilon} = 0}=  & -\left\{
\left(\frac{\phi_{rrr}^{\ell k}}{(r^{\ell k})^3} -
\frac{3\phi_{rr}^{\ell k}}{(r^{\ell k})^4} + \frac{3\phi_{r}^{\ell k}}{(r^{\ell k})^5}\right)
r^{\ell k}_\alpha r^{\ell k}_\beta r^{\ell k}_\nu r^{\ell k}_\chi \right.  \\
&+\left(\frac{\phi_{rr}^{\ell k}}{(r^{\ell k})^2} -\frac{\phi_r^{\ell k}}{(r^{\ell k})^3}\right)
\left(
\delta_{\alpha\chi}r^{\ell k}_\beta r^{\ell k}_\nu \!+\!
\delta_{\beta\chi}r^{\ell k}_\alpha r^{\ell k}_\nu \!+\!
\delta_{\nu\chi}r^{\ell k}_\alpha r^{\ell k}_\beta \!+\!
\delta_{\nu\alpha}r^{\ell k}_\beta r^{\ell k}_\chi \!+\!
\delta_{\nu\beta}r^{\ell k}_\alpha r^{\ell k}_\chi
\right) \\
&+\frac{\phi_r^{\ell k}}{r^{\ell k}}\left(
\delta_{\nu\alpha}\delta_{\beta\chi} +
\delta_{\nu\beta}\delta_{\alpha\chi}
\right)
\left. \vphantom{\frac{\sum}{}}\right\}\ ,
\end{split}\end{equation}
and the diagonal part is just
\begin{equation}
\left.\frac{\partial^3 U}
{\partial x^\ell_\chi \partial x^\ell_\nu \partial \epsilon_{\alpha\beta}}
\right|_{\mathBold{\epsilon} = 0}
= -\sum_k \frac{\partial^3 U}
{\partial x^\ell_\chi \partial x^k_\nu \partial \epsilon_{\alpha\beta}} \ .
\end{equation}
\end{widetext}


\begin{thebibliography}{99}
\bibitem{08ELF}
E. Bouchbinder, A. Livne, J. Fineberg, Phys. Rev. Lett. 101, 264302 (2008).

\bibitem{10LBSF}
A. Livne, E. Bouchbinder, I. Svetlizky, J. Fineberg, Science 327, 1359 (2010).

\bibitem{10KLLP}
S. Karmakar, A. Lemaitre, E. Lerner, I. Procaccia, ``Predicting Plasticity in Amorphous Solids", Phys. Rev. Lett, submitted. ArXiv: arXiv:1002.3487.

\bibitem{10KLP}
S. Karmakar, E. Lerner and I. Procaccia, ``Plasticity-Induced Anisotropy in Amorphous Solids: the Bauschinger Effect". Phys Rev E submitted. Also: ArXiv: arXiv:0910.4281.

 \bibitem{04ML}
C. Maloney and A. Lemaitre, Phys. Rev. Lett. {\bf 93},  195501  (2004).

\bibitem{06ML}
A. Lemaitre and C. Maloney, J. Stat. Phys. {\bf 123},  415  (2006).

\bibitem{09LPa}
E. Lerner and I. Procaccia, Phys. Rev. E {\bf 79}, 066109 (2009).

\bibitem{09LPb}
E. Lerner and I. Procaccia, Phys, Rev. E, 80, 026128 (2009).

\bibitem{09HKLP}
H.G.E. Hentschel, S. Karmakar, E. Lerner, and I. Procaccia, Phys. Rev. Lett. {\bf 104}, 025501 (2009).

\bibitem{89Lutsko}
J. F. Lutsko, J. Appl. Phys. {\bf 65}, 2991 (1989).

\bibitem{90BCH}
J.A. Bannantine, J.J. Comer and J.L. Handrock,  {\em Fundamentals of Metal Fatigue Analysis},  (Prentice-Hall, 1990).


\end{thebibliography}
\end{document}